\documentclass[journal]{IEEEtran}
\usepackage{cite}
\usepackage{multirow}
\usepackage{graphicx}
\usepackage{amssymb,amsmath,times,wasysym}
\usepackage{pb-diagram}
\usepackage{psfrag,balance}
\usepackage{srcltx,verbatim,color} 
\usepackage[caption=false]{subfig} 
\newcounter{mycounter}
\usepackage[margin=0.75in]{geometry}
\usepackage{algorithmic}
\usepackage{algorithm}
\usepackage[ansinew]{inputenc} 
\usepackage{fixltx2e}
\usepackage{multicol}
\usepackage{tabularx}

\newcounter{appendx}

\newcommand{\E}[2][]{\mathbb{E}_{#1}\!\left[{#2}\right]}

\newcommand{\Var}[1]{\mathbb{V} \mathrm{ar}\!\left[{#1}\right]}

\newcommand{\e}[1]{\text{exp}\!\left({#1}\right)}

\renewcommand{\log}[2][]{\mathrm{log}_{#1}\left(#2\right)}

\title{\Huge{Exploiting  Cell-Free Massive MIMO for Enabling Simultaneous Wireless Information and Power Transfer} \vspace{-0mm}}

\author{{Diluka Loku Galappaththige, \IEEEmembership{Student Member, IEEE}, Rajan Shrestha,     and  Gayan Amarasuriya  Aruma Baduge, \IEEEmembership{Senior Member, IEEE}  \vspace{-0mm} }
	
\thanks{Authors are with the Department of Electrical  and Computer Engineering, Southern Illinois University, Carbondale, IL, USA 62901, Email: \{diluka.lg, rajan.shrestha, gayan.baduge\}@siu.edu.  This work in part has been presented  at Global Communications Conference (Globecom), 2018, Abu Dhabi, UAE \cite{Rajan2018}. 
	}\vspace{-10mm}}

\begin{document}
\bstctlcite{IEEEexample:BSTcontrol}
\vspace{-10mm}
\maketitle

\begin{abstract}\\

The   performance  of simultaneous wireless information and power transfer (SWIPT) in downlink (DL) cell-free massive multiple-input multiple-output (MIMO) is investigated.   Tight approximations to the DL harvested energy and the DL/uplink (UL) achievable rates   are  derived    for two practical channel state information (CSI) cases by using a non-linear energy harvesting model for time-switching and power-splitting protocols.
Max-min fairness-based transmit power control policies are employed to mitigate the deleterious near-far effects caused by  distributed transmissions/receptions in cell-free massive MIMO. 
The achievable common DL energy-rate trade-off is derived, and thereby, it is shown that the proposed max-min power control   guarantees user-fairness regardless of near-far effects in terms of both harvested energy and achievable rate.   The benefits of user estimated DL  CSI to boost the SWIPT performance are explored. 
These performance metrics are compared against those of the conventional co-located massive MIMO, and thereby, it is revealed that  the  reduction of path-losses and lower average transmit powers offered by cell-free massive MIMO can be exploited to boost the energy-rate trade-off of SWIPT at the expense of increased backhaul requirements.

\end{abstract}


\linespread{1.2}


\section{Introduction}\label{sec:introduction}

The concept of wireless power transfer (WPT) was  conceived by Nikola Tesla in the 1890s. It was 
originally intended for long-distance transmission of electrical energy from a power source to an electrical load without wired interconnections. Recently, the short-range WPT technology was rejuvenated for prolonging the battery-life of energy-constrained user nodes such as low-complexity  sensors and  Internet-of-Things (IoTs) \cite{Lin2017}. In the downlink (DL)   WPT, the energy is beamformed towards   the users, and  this harvest energy is typically stored or used to charge batteries. Then,  a predefined fraction of the DL harvested energy can be used for uplink (UL)  transmissions  \cite{Ng2013,Yao2017}. 

By exploiting the broadcast nature of radio frequency (RF) wave propagation,  the concept of simultaneous wireless information and power transfer (SWIPT) has  been envisioned  recently  \cite{Varshney2008,Zhang2013,Visser2013}. SWIPT techniques can be leveraged to prolong the battery-life or to serve as an alternative power source  of energy-constrained, low-power wireless nodes in emerging IoTs \cite{Rajan2018,Varshney2008,Zhang2013,Ding2015,Amarasuriya2016,Ponnimbaduge2018,Nasir2013,Krikidis2014}.  Deviating from separated information/energy receiver architectures \cite{Varshney2008,Visser2013}, two low-complexity  co-located receiver structures, namely (i) the power-splitting (PS) receiver and (ii) the time-switching (TS) receiver, have been   proposed in \cite{Zhang2013,Nasir2013} for  enabling SWIPT in wireless systems. In PS receiver, the RF power captured by the rectenna  is split based on a PS ratio  by  a power splitter and directed toward the information decoding and energy harvesting \cite{Zhang2013}. In TS receiver,      the transmission duration is split into two orthogonal time-slots, which are then used for information and energy  transmissions \cite{Zhang2013}.

Despite its promise projected by early exploratory research with idealized system models \cite{Varshney2008,Zhang2013}, the main inhibiting factors of enabling SWIPT in practical wireless networks include (i) inherent severe end-to-end path-loss   \cite{Ding2015}, (ii) low radio frequency to direct current (RF-to-DC) conversion efficiency of rectennas \cite{Ding2015,Amarasuriya2016}, and (iii) non-linear operating characteristics of energy harvesting circuitry \cite{Hagerty2014,Wang2017S,Wang2017Y}. Nonetheless,  favorable propagation,  channel hardening, and aggressive spatial multiplexing gains    rendered by massive multiple-input   multiple-output (MIMO) can be exploited to boost the performance gains of SWIPT \cite{Rajan2018,Ding2015,Amarasuriya2016}.    

Massive MIMO is originally conceived to serve many users in the same time-frequency resource element by exploiting aggressive spatial multiplexing gains offered by a very large co-located antenna array at a base-station (BS) \cite{Marzetta2010}. Time-division duplexing (TDD) based co-located massive MIMO in sub-6\,GHz frequency band has already been commercialized in the United States by Sprint \cite{Sprint}, and it is one of the   key enabling technologies of the fifth-generation (5G) standard \cite{release15}. Co-located massive MIMO is primarily intended for conventional cellular type deployments.    

Recently, a cell-free massive MIMO architecture has been envisioned for wireless standards beyond 5G \cite{Ngo2015}. In cell-free massive MIMO,   a large number of geographically distributed antennas or access-points (APs) simultaneously serves many users   in the same time-frequency resource element in the TDD mode of operation. All APs are connected to   a  central processing unit (CPU) via a conventional backhaul network \cite{Ngo2017} or through a fronthaul network \cite{Interdonato2019} where the CPU is a part of an emerging cloud-radio access network (C-RAN) \cite{Yang2015a}. The main difference between the cell-free massive MIMO and the well-established concepts of network MIMO \cite{Karakayali2006}, coordinated-multipoint  (CoMP) \cite{Irmer2011}, and distributed antenna systems  (DAS) \cite{Choi2007} is that   the perspectives of cell-boundaries are not considered in the former because  all APs, which   are distributed across a given geographical area, cooperatively/coherently serve all user nodes via spatial multiplexing techniques \cite{Ngo2017}.  
The TDD-based operation of cell-free massive MIMO enables local estimation of UL  channel state estimation (CSI) at each AP \cite{Ngo2017}, and thus, the overhead of CSI exchange among APs can be mitigated.   The DL payload data and transmit power allocation coefficients of the user nodes are sent by the CPU to all APs in which  a simple  maximal ratio transmission (MRT) based precoder can be used for DL transmissions \cite{Ngo2017}. The APs forward the data received from the user nodes in the UL towards the CPU through the fronthaul/backhaul network. Then, a maximal ratio combining (MRC) based joint detection of UL user signals can be conducted at the CPU by adopting the statistical DL CSI whenever DL channel estimation at the user nodes is  utilized to minimize the channel estimation overhead \cite{Ngo2017}.   
The benefits of cell-free massive MIMO include high probability of coverage for user nodes with better average channel conditions with  average reductions of path-losses, large macro-diversity gains, reduced average transmit power requirements, and large  spectral/energy efficiencies \cite{Ngo2017}.

\subsection{SWIPT with co-located massive MIMO}

In \cite{Chen2013}, the maximization of energy efficiency, while satisfying quality-of-service   requirement, is investigated for energy beamforming based  SWIPT in large-scale MIMO systems.  
In \cite{Yang2015}, SWIPT for hybrid data/energy users is studied in the context of
massive MIMO   by proposing a max-min throughput optimization technique based on optimal allocation of time-slot durations and transmit power.   In \cite{Zhao2016},  the performance of SWIPT in DL massive MIMO is investigated by maximizing the   minimum harvested energy among the energy-users  subject to a  minimum  achievable rate for the data-users.  The performance bounds and power control coefficients of \cite{Yang2015} and \cite{Zhao2016} are valid only for the asymptotically large BS antenna regime.
In \cite{Kudathanthirige2019},  the impact of imperfectly estimated CSI on SWIPT performance in massive MIMO is studied by quantifying the max-min optimal energy-rate trade-off.     
In \cite{Zhu2016}, efficient user association techniques for wireless energy transfer in massive MIMO heterogeneous networks are proposed. 
In \cite{Kudathanthirige2019b,Shrestha2019}, the feasibility of adopting  dual-hop massive MIMO relay networks in boosting the performance of SWIPT is investigated.  
In \cite{Amarasuriya2016}, the performance metrics of   SWIPT  in multi-cell  massive MIMO multi-way relay networks are derived in the presence of erroneously estimated CSI. 

\subsection{Motivation}

Although SWIPT  has been extensively adopted for co-located massive MIMO \cite{Chen2013,Yang2015,Amarasuriya2016,Zhao2016,Kudathanthirige2019}, 
to the best of our knowledge, practically viable  performance bounds of SWIPT   for cell-free massive MIMO have not yet been investigated in the open literature except for the initial idea   in \cite{Yuan2015}.  Thus, our paper fills this gap by establishing practically achievable performance bounds of  SWIPT in cell-free massive MIMO with   estimated CSI. The proposed system model can reap benefits of distributed transmission/reception in cell-free massive MIMO to overcome the challenge of low harvested energy levels observed in SWIPT primarily owing to  inherent severe path-losses.  Since the APs are  distributed over a given large geographical area, cell-free massive MIMO leverages macro-diversity benefits to mitigate shadow fading more efficiently than the co-located counterpart. Because the users   tend to be located more  closer to the APs, SWIPT with cell-free massive MIMO  has a potential of offering  substantially higher  coverage probability, while minimizing the throughput/energy outage probabilities than the co-located massive MIMO.    Cell-free massive MIMO has been shown to be more robust against the detrimental correlated small/large-scale fading than the co-located counterpart \cite{Ngo2017}. 
Thus, the cell-free massive MIMO can   boost the performance of SWIPT, and  the underlying practical potential motivates our work. 

\subsection{Our contribution}

In this paper, we leverage the  benefits of cell-free massive MIMO to boost the performance of a hybrid  SWIPT/WPT system model. The information and energy are transfered in the DL by adopting SWIPT-enabled cell-free massive MIMO, and then, the DL harvested energy is used for UL user transmissions.   Tight approximations to the  achievable  rates, harvested energy, and fundamental rate-energy trade-offs of SWIPT-enabled cell-free massive MIMO are derived  for two PS and TS receiver structures.  These performance metrics capture     non-linear operating characteristics of the energy harvesting circuitry \cite{Wang2017S}.

Our performance metrics are  categorized  by considering the availability of statistical  CSI and estimated DL CSI at the users. In the former case, the   users s rely on statistical CSI for signal decoding, while in the latter case, the estimated DL CSI is used at the users by virtue of DL pilots beamformed by the APs. The energy-rate trade-offs of these two  CSI cases are quantified, and our analysis captures the deleterious effects of residual interference caused by  beamforming/detection uncertainties at the APs/users
owing to imperfect/partial CSI.  

 This paper technically contributes to the mitigation of detrimental near-far effects on   SWIPT performance metrics of a cell-free massive MIMO set-up. The achievable rate-energy trade-offs of   the    
max-min fairness optimal transmit power control are derived to maximize the minimum achievable user rate and the harvested energy for both TS and PS receiver structures. By adopting the corresponding optimal transmit power control coefficients, the common user rates and harvested energies  are derived. Then, the max-min fairness optimal energy-rate trade-offs are  quantified and compared against those of  the uniform power control. Thereby, it is revealed that the proposed power control policies are able to mitigate the near-far effects of SWIPT-enabled cell-free massive MIMO.

Our  analysis is  then used to quantify the performance gains of SWIPT-enabled cell-free massive MIMO over  the conventional  co-located counterpart. Consequently, it is shown that the former set-up outperforms the latter mainly owing to   inherent macro-diversity gains and lesser average path-losses caused by distributed APs.  However,  these gains are achieved by cell-free set-up   at the expense of increased backhaul/fronthaul requirements over the co-located counterpart. 

Finally, the feasibility of utilizing the  energy  harvested during the DL SWIPT at the users for UL transmission  is investigated by deriving the UL achievable rates. Our analysis reveals that the UL user rate gains for this specific case primarily rely on efficient  DL/UL transmit power control. This is because the harvested energy via DL SWIPT depends on DL power control at the APs, while the UL  rates depend on how the DL harvested energy is allocated for UL transmit power and battery storage at the users. It is advocated that the  transmit power must be jointly controlled to optimize the DL harvested energy and UL achievable rates. The deleterious  near-far effects on the energy-rate trade-off for this specific case have shown to be more stringent with the cell-free set-up than the co-located counterpart. Nevertheless, it is also revealed that the proposed max-min power control policies can jointly mitigate near-far effects not only on both DL harvested energy but also DL/UL achievable rates.

\subsection{Difference relative to the existing literature:}    
Even though there exists a handful of research on SWIPT in co-located massive MIMO  \cite{Chen2013,Yang2015,Zhao2016,Kudathanthirige2019,Zhu2016,Kudathanthirige2019b,Shrestha2019,Amarasuriya2016}, the  fundamental  performance metrics  have not yet been established in the context of SWIPT-enabled cell-free massive MIMO. 
It has been already shown that the data rate performance of cell-free massive MIMO    is  fundamentally different from that of  the co-located counterpart \cite{Ngo2017}.
Thus, the performance bounds of SWIPT in cell-free massive MIMO are also expected to be considerably different from those established for co-located  counterpart  \cite{Kudathanthirige2019b,Shrestha2019,Kudathanthirige2019,Yang2015,Zhao2016,Zhu2016,Amarasuriya2016}. This is because   not only users but also APs are geographically distributed, and hence, the harvested energy and   rates are severely hindered by the near-far effects. In \cite{Kudathanthirige2019},  
max-min transmit power control has been adopted to separately mitigate  detrimental    near-far effects on  the   rates and harvested energy in SWIPT-enabled co-located massive MIMO. However, in a cell-free set-up, near-far effects on both  the rate and harvested energy must be jointly mitigated. To this end, we investigate max-min  transmit power control policies to optimize the   energy-rate trade-off, while guaranteeing the user-fairness.    The distinct long-term channel coefficients between the distributed APs and the users   ensure  that our  max-min  power control solutions are  different from those appeared for the co-located counterpart, where the large-scale channel coefficient between a given user and all antennas  is the same. 

All related prior works on   SWIPT-enabled massive MIMO     \cite{Chen2013,Yang2015,Zhao2016,Kudathanthirige2019,Zhu2016,Amarasuriya2016,Kudathanthirige2019b,Shrestha2019} utilize   linear energy harvesting models at the users. Nonetheless, our analysis deviates from this ideal model by adopting a practically viable  SWIPT  model proposed by  \cite{Wang2017S}, which can capture  non-linear characteristics of energy harvesting circuitry. Moreover, by adopting a  hybrid TS/PS protocol from \cite{Atapattu2016}, we generalize our performance analysis for both TS and PS protocols. 

Although the impact of imperfectly estimated UL channels has been investigated to design precoders for  SWIPT-enabled massive MIMO \cite{Kudathanthirige2019b,Shrestha2019}, the users have been assumed to rely on statistical CSI. This assumption can be justified for co-located massive MIMO because the instantaneous channel coefficients can  be approximated tightly by their average counterparts when the number of co-located antennas grows without bound \cite{Marzetta2016_Book}. However, the  same justification may not be valid for a cell-free set-up in which the APs are  geographically distributed  \cite{Ngo2016}. To this end, distinctively from the  related prior research \cite{Chen2013,Yang2015,Zhao2016,Kudathanthirige2019,Zhu2016,Amarasuriya2016,Kudathanthirige2019b,Shrestha2019}, we investigate  the impact of user estimated DL CSI, which is acquired through beamforming DL pilots towards the users, and thereby deriving the SWIPT performance metrics.   To the best of our knowledge, the proposed DL/UL max-min fairness optimal energy-rate trade-offs of DL and UL transmission in SWIPT have not yet been reported in the context of cell-free massive MIMO.  
      
\noindent \textbf{Notation:}      $\mathbf z^{H}$ and $\|\mathbf z\|$  denote the conjugate-transpose and Euclidean norm of $\mathbf z$, respectively. The conjugate of $z$ is denoted by $z^*$, and   $z\sim \mathcal {CN}(\cdot,\cdot)$ denotes that $z$ is a complex-valued circularly symmetric Gaussian   random variable.   $\E{\cdot}$ and $\Var{\cdot}$ are the expectation and variance, respectively.

\begin{figure}[!t]\centering \vspace{-0mm}
	\def\svgwidth{180pt} 
	\fontsize{7}{4}\selectfont 
	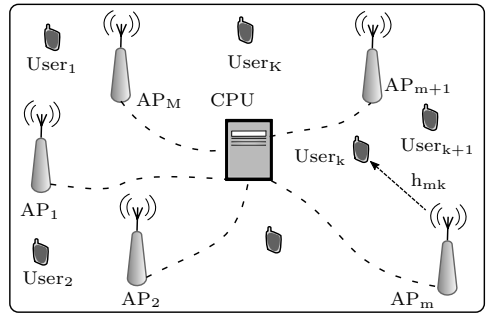 \vspace{-3mm}
	\caption{A cell-free massive MIMO system.}\vspace{-0mm} \label{fig:system_Model}
\end{figure}

\section{System, channel and signal models  }\label{sec:system_model}

\subsection{System   model}\label{sec:system_and_channel}

We consider a  TDD-based cell-free massive MIMO system  in which $M$ single-antenna APs serve $K$ single-antenna user nodes  (see Fig. \ref{fig:system_Model}). The user nodes are equipped with both information and energy receivers. The energy harvesting model at each user node is generalized  by adopting both   TS and PS receiver structures/protocols. 
The channel coherence interval ($\tau_{c}$) is divided into three main orthogonal time-slots and used for  pilot,  DL, and  UL  transmissions. Depending on the CSI case\footnote{The pilot transmission and channel estimation aspects are discussed in Sections \ref{sec:Channel_state_information_acquisition} and \ref{sec:Impact_DL_pilots}.}, the pilot phase is used for either  only UL and  or both UL/DL pilot transmissions  (see Fig. \ref{fig:time_fram}a and Fig. \ref{fig:time_fram}b).

\subsubsection{SWIPT in DL transmission}\label{sec:SWIPT_in_DL}
The SWIPT  is performed in DL transmission phase by using  TS protocol,  PS protocol or both protocols. Thus, 	
the DL time-slot  is further divided into two portions based on  the TS/PS factors $(\alpha$/$\theta)$ to receive payload data and harvest the energy. Specifically, $\alpha \tau_d$ portion of the DL time-slot is dedicated for DL power transfer via TS protocol, while the reminder portion of  length $(1-\alpha)\tau_d$ is used for simultaneous DL information  transfer. If the PS protocol is adopted, then   the received signal power is split into two  streams during $(1-\alpha)\tau_d$ via a power splitter having a PS ratio of $\theta:1-\theta$  and passed through the energy harvester and the information decoder, respectively. 

\subsubsection{UL transmission}\label{sec:UL_tx_1}
\begin{figure}[!t]\centering \vspace{-0mm}
	\def\svgwidth{250pt} 
	\fontsize{7}{4}\selectfont 
	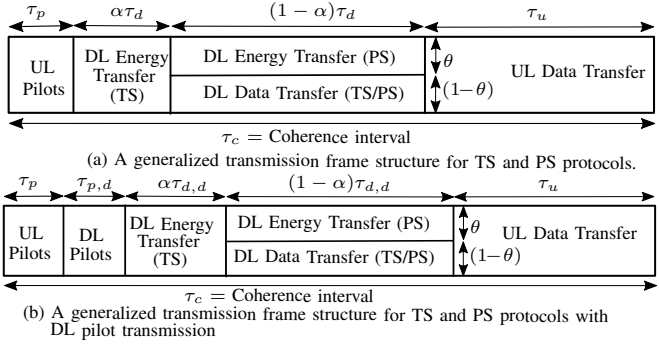 \vspace{-3mm}
	\caption{A generalized transmission frame structure for TS and PS protocols with/without DL pilots.}\vspace{-0mm} \label{fig:time_fram}
\end{figure}

The final time-slot $(\tau_u)$ of the coherence interval is allocated for the UL\footnote{The performance analysis corresponding to joint DL/UL transmission within the same coherence interval is discussed in Section \ref{sec:UL_tx}.} user data transmission towards APs.  During this UL time-slot, each user node allocates a fraction of its harvested energy in the DL   phase  for UL data transmission   (see Fig. \ref{fig:time_fram}a). 

\subsubsection{Backhaul/fronthaul requirement}\label{sec:backhaul}
In cell-free massive MIMO, all APs must be connected to a CPU  either via  a conventional backhaul network \cite{Ngo2017} or by using an emerging fronthaul networks in C-RAN \cite{Interdonato2019}. Since the UL channels are estimated locally at each AP via user pilots and a local MRT precoder is employed at each AP,  no CSI is exchanged among APs.  This considerably reduces the overhead of cell-free systems compared to network MIMO  and CoMP. However, DL/UL data must be transmitted to/from the CPU towards APs. Thus, there exists a fundamental trade-off between performance gains and backhaul/fronthaul requirements.

\subsection{Channel model}\label{sec:channel_model}
The channel coefficient between the $m$th AP and the $k$th user is denoted by $h_{mk}$, and the    channel is modeled as 
	\begin{eqnarray}\label{eqn:Channel_coefficient}
	{h}_{mk}  =   \zeta_{mk}^{1/2} \tilde{h}_{mk},
	\end{eqnarray}
 where $\tilde{h}_{mk} \sim \mathcal {CN}(0,1)$ captures     small-scale/Rayleigh fading
and stays fixed during the coherence interval \cite{Marzetta2016_Book}. Moreover,  $\zeta_{mk}$ captures the large-scale fading effects, including path-loss  and log-normal shadowing, between the $m$th AP and the $k$th user. It is typically assumed that    these large-scale fading coefficients   are known a-priori  as they   change  slowly and remain fixed for thousands of  coherence interval \cite{Marzetta2016_Book,Ngo2017}. 

\subsubsection{Uplink channel estimation}\label{sec:Channel_state_information_acquisition}

The UL channels are estimated locally at   APs by using user pilots \cite{Ngo2017}. During the UL  time-slot, all $K$ users send pilots  of length $\tau_{p}$ to all APs. The pilot   transmitted by the $k$th user is denoted by  $\boldsymbol{\phi}_k \in \mathbb C^{ 1 \times \tau_{p}}$, where $k \in \{1,\cdots,K\}$. We assume that   these pilots   are mutually orthogonal;  $\boldsymbol{\phi}_k^H \boldsymbol{\phi}_{k'} = 0 $ for $k \neq k'$ and $\| \boldsymbol{\phi}_k \|^2 = 1$. Thus,  the pilot length should satisfy $\tau_{p} \geq K$.
The pilot signal received at the $m$th AP is given by  \cite{Marzetta2016_Book,Ngo2017}
\begin{eqnarray}\label{eqn:Rx_pilot_vector}
	\mathbf{y}_{pm}' = \sqrt{\tau_{p} P_{p}} \sum_{k=1}^{K} {h}_{mk} \boldsymbol{\phi}_k + \mathbf n_m',
\end{eqnarray} 
 where $P_p$ is the pilot transmit power of each user, $\mathbf {n}_m'$ is an additive white Gaussian noise (AWGN) vector at the $m$th AP, and elements of $\mathbf {n}_m'$ are independent and identically distributed  (i.i.d.)  $\mathcal {CN}(0,1)$ random variables.
By projecting   \eqref{eqn:Rx_pilot_vector} onto $\boldsymbol{\phi}_{k}^H$, a sufficient statistic that can be used to  estimate  $h_{mk}$ locally at the $m$th AP can be written as \cite{Marzetta2016_Book,Ngo2017}
\begin{eqnarray}\label{eqn:pilot_estimate}
	{y}_{pmk} = \boldsymbol{\phi}_{k}^H \mathbf{y}_{pm}' = \sqrt{\tau_{p} P_{p}}  {h}_{mk}  +  n_m,
\end{eqnarray}
 where $n_m = \boldsymbol{\phi}_{k}^H \mathbf n_m'\sim \mathcal {CN}(0,1)$ because $\boldsymbol{\phi}_{k}$ is a unitary vector.  
The   minimum mean square error (MMSE)   estimate of $h_{mk}$ can be derived as \cite{Kay1993}
\begin{eqnarray}\label{eqn:estimate_of_h_mk}
	\hat {h}_{mk} = \frac{\E {y_{pmk}^* h_{mk}}}{\E {|y_{pmk}|^2}}  y_{pmk} = c_{mk} y_{pmk} ,
\end{eqnarray} 
 where $c_{mk}$ is given by
\begin{eqnarray}\label{eqn:c_mk}
	c_{mk} = \frac{\sqrt{\tau_{p} P_p} \zeta_{mk}}{\tau_{p} P_p \zeta_{mk} + 1}.
\end{eqnarray} 
 Due to  channel reciprocity property of TDD,  APs utilize the same locally estimated $\hat {h}_{mk}$ as DL CSI to construct its precoder  \cite{Marzetta2016_Book}. The true channel can be represented as $h_{mk} = \hat{h}_{mk} + \epsilon_{mk}$, where $\epsilon_{mk}$ is the   estimation error, which is independent of the estimate yielded from the orthogonality property of MMSE criterion \cite{Kay1993}.

\subsection{Non-linear energy harvesting model}\label{sec:Non_linear_energy_harvesting_ model}

In practice, the harvested energy is a non-linear function of the received power at the rectenna. Such a  non-linear energy harvesting circuitry can be modeled as \cite{Wang2017S}
\begin{eqnarray}\label{eqn:EH_model} 
	\Psi_{EH} \left(P_{R}\right)  &=&  \left[\frac{\lambda}{1-\nu} \left( \frac{1}{1+ \e{-\mu \left(P_R - \omega\right)}}- \nu \right) \right]^+ \nonumber \\
	 &=& \left[ \frac{\lambda \left(1 - \e{-\mu P_R}\right)}{1 + \e{-\mu \left(P_R - \omega\right)}}\right]^+,
\end{eqnarray}
 where $\Psi_{EH} \left(P_{R}\right)$ is a non-linear function of $P_R$, and represents  the instantaneous harvested energy. Moreover,  $\nu = 1/\left(1+\e{\mu \omega}\right)$ and $\left[z\right]^+ = \max \left(0,z\right)$. The   non-linearities  of the energy harvesting circuitry are modeled via $\lambda$, $\mu$ and $\omega$.   Based on \cite{Wang2017S}, these parameters are set to  $\lambda = 20$\,mW, $\mu = 6400$\,$/\mu$W, and $\omega = 2.9$\,$\mu$W.  

\subsection{Signal model}\label{sec:Signal_model}

\subsubsection{Signal model for DL transmission}\label{sec:Signal_model_DL}
In DL transmission, APs employ conjugate precoding, which is designed based on the locally estimated channels (\ref{eqn:estimate_of_h_mk}). The  signal transmitted by the $m$th AP can be written as
\begin{eqnarray}\label{eqn:mth_Tx_signal} 
	x_{m}  =  \sqrt{P_d} \sum_{k=1}^{K} {{\eta_{mk}^{1/2}}  \hat {h}_{mk}^* q_{k}} , 
\end{eqnarray}
 where $q_k$ is the symbol intended for the $k$th user  satisfying $\E{|q_k|^2} = 1$. In \eqref{eqn:mth_Tx_signal},  $P_d$ is the maximum allowable transmit power at each AP, and $\eta_{mk}$ is the   power allocation to  the $k$th user symbol at the $m$th AP. Then, the transmit power constraint   of the $m$th AP can be defined as $\E{|x_m|^2} \leq P_d$. By substituting (\ref{eqn:mth_Tx_signal}) into this constraint and applying several mathematical  manipulations, it can be  rewritten as
\begin{eqnarray}\label{eqn:Tx_signal_power}
\sum_{k=1}^{K} \eta_{mk} \E{|\hat{h}_{mk}^*|^2} \leq 1 \quad \Rightarrow \quad  \sum_{k=1}^{K} \eta_{mk} \rho_{mk} \leq 1,
\end{eqnarray} 
 where  $\rho_{mk} \triangleq \E {|\hat{h}_{mk}|^2} = \sqrt{\tau_{p} P_p} c_{mk} \zeta_{mk}$.
The signal  received at the $k$th user from all $M$ APs can be written as 
\begin{eqnarray}\label{eqn:kth_Rx_signal}
	r_k = \sum_{m=1}^{M} {h_{mk} {x}_{m}} + n_{k},
\end{eqnarray}	
  where $n_{k}\sim \mathcal{CN}(0,1)$ is an AWGN at the $k$th user.  The signal in (\ref{eqn:kth_Rx_signal}) can be rearranged as
\begin{eqnarray}\label{eqn:rearranged_kth_Rx_signal}
	r_k &=& \sqrt{P_d} \sum_{m=1}^{M} {\eta_{mk}^{1/2}}  h_{mk} \hat{h}_{mk}^* q_k \nonumber\\
	&&+ \sqrt{P_d} \sum_{m=1}^{M} \sum_{i \neq k}^{K} {\eta_{mi}^{1/2}}  h_{mk} \hat{h}_{mi}^* q_i + n_{k},
\end{eqnarray}	 
  where the first term represents the desired signal component, while the second term captures the inter-user interference yielded from beamforming uncertainty of conjugate precoding with imperfectly estimated  CSI at the APs.

\section{Preliminary analysis}\label{sec:Performance_analysis}

\subsection{Instantaneous Harvested Energy}\label{sec:Instantaneous_Harvested_Energy}

During the first portion of DL time-slot having a length of $\alpha \tau_d$, the APs transfer power in the DL and users harvest energy based on the TS protocol (see Fig. \ref{fig:time_fram}a). 
During the second portion  having a length of $(1-\alpha) \tau_d$, users harvest energy based on the PS protocol. 
The total harvested  energy for TS and PS protocols at the $k$th user can be written as
\begin{eqnarray}\label{eqn:instantaneous_HE}
	E_k = \alpha \tau_d \Psi_{EH} \left(P_k\right)  +  (1-\alpha) \tau_d \Psi_{EH} \left(\theta P_k\right),
\end{eqnarray} 
   where 	$\Psi_{EH}(\cdot)$ is the non-linear function   in (\ref{eqn:EH_model}), and $P_k$ is the incident/received power at the rectenna of the $k$th user. By using (\ref{eqn:kth_Rx_signal}), $P_k$ can be defined as 
\begin{eqnarray}\label{eqn:instant_Rx_power_k_user}
	P_k = P_d \left| \sum_{m=1}^{M} \sum_{i=1}^{K} {\eta_{mi}^{1/2}} h_{mk} \hat{h}_{mi}^*  q_i \right|^2.
\end{eqnarray} 
\textbf{\textit{Remark 1:}} The harvested energy   (\ref{eqn:instantaneous_HE}) can be used to investigate the performance of both TS and PS protocols. By letting $\theta=0$ and $\alpha\neq 0$, the harvested energy yielded by the TS protocol can be computed. If $\alpha=0$ and $\theta\neq 0$,  then (\ref{eqn:instantaneous_HE}) provides the harvested energy  of the PS protocol. When $\alpha\neq 0$ and $\theta \neq 0$, (\ref{eqn:instantaneous_HE}) can be used to quantify the performance of a hybrid TS/PS protocol. The same argument applies to the achievable user rate, which is presented in (\ref{eqn:kth_US_rate}). 

\setcounter{mycounter}{\value{equation}}
\begin{figure*}[!t] 
	\vspace{-7mm}
	\addtocounter{equation}{3}
		\begin{eqnarray}\label{eqn:k_th_Us_WC_Gaussian_Rx_signal}
		r_k &=& \underbrace{\sqrt{\tilde{P_d}} \E {\sum\nolimits_{m=1}^{M} \eta_{mk}^{1/2} h_{mk} \hat{h}_{mk}^* q_k }}_{\text{desired signal}} + \underbrace{\sqrt{\tilde{P_d}}  \left(\sum\nolimits_{m=1}^{M} \eta_{mk}^{1/2} h_{mk} \hat{h}_{mk}^* q_k  - \E{\sum\nolimits_{m=1}^{M} \eta_{mk}^{1/2} h_{mk} \hat{h}_{mk}^* q_k } \right) }_{\text{detection uncertainty}} \nonumber \\
		&&+ \underbrace{\sqrt{\tilde{P_d}} \sum\nolimits_{m=1}^{M} \sum\nolimits_{i \neq k}^{K} \eta_{mi}^{1/2} h_{mk} \hat{h}_{mi}^* q_i }_{\text{Inter-user interference}} + \underbrace{n_k}_{\text{AWGN}}
		\end{eqnarray}
	\vspace{-9mm}
\end{figure*}
\setcounter{equation}{\value{mycounter}}
\setcounter{mycounter}{\value{equation}}
\begin{figure*}[!t] 
	\addtocounter{equation}{4}
	\begin{eqnarray}\label{eqn:SINR_kth_US}
	\!\!\!\!\! \!\!\!\!\!
	\gamma_k = \frac{\tilde{P_d} \left| \E{\sum_{m=1}^{M} \eta_{mk}^{1/2} h_{mk} \hat{h}_{mk}^* } \right|^2}{\tilde{P_d} \Var{\sum_{m=1}^{M}\eta_{mk}^{1/2} h_{mk} \hat{h}_{mk}^* } +  \tilde{P_d}  \E{\left|\sum_{m=1}^{M} \sum_{i \neq k}^{K} \eta_{mi}^{1/2} h_{mk} \hat{h}_{mi}^* \right|^2 } + \E{\left|n_k\right|^2} },
	\end{eqnarray}
	\vspace{-2mm}
	\hrulefill
\end{figure*}
\setcounter{equation}{\value{mycounter}}

\subsection{Average Harvested Energy}\label{sec:Average Harvested Energy}

The average harvested energy at the $k$th user is given by  
\begin{eqnarray}\label{eqn:average_HE}
	\bar{E}_k = \alpha \tau_d \E{\Psi_{EH} \left(P_k\right)}  +  (1-\alpha) \tau_d \E{\Psi_{EH} \left(\theta P_k\right)}.
\end{eqnarray} 
    The exact closed-form evaluation of \eqref{eqn:average_HE} appears mathematically intractable. To obtain  useful insights, a tight upper bound for \eqref{eqn:average_HE} is  derived by using Jensen's inequality as 
 \begin{eqnarray}\label{eqn:avg_HE_up_bound}
	\!\!\!\!\!\!\bar{E}_k \!\leq \! \bar{E}_k^{ub} \!= \! \alpha \tau_d \Psi_{EH} \left( \E{P_k} \right) \!+\!  (1 \!-\!\alpha) \tau_d \Psi_{EH} \left(\theta \E{ P_k}\right)\!,
\end{eqnarray}  
   where $\E{P_k} = 	\bar{P}_k$ is derived    as  (see Appendix \ref{app:Appendix2})
\begin{eqnarray}\label{eqn:avg_power_kth_Us}
\E{P_k} =   P_d \sum_{m=1}^{M} \eta_{mk} \rho_{mk}^2 +  P_d \sum_{m=1}^{M} \sum_{i=1}^{K} \eta_{mi} \rho_{mi} \zeta_{mk}.
\end{eqnarray}

\subsection{Achievable DL rate}\label{sec:Achievable_DL_rate}

 The users are unaware of instantaneous channel coefficients with  no DL pilots, and thus, they    must rely on the statistical CSI   for the signal decoding  \cite{Marzetta2016_Book}. The  signal received at the $k$th user \eqref{eqn:rearranged_kth_Rx_signal} can be re-arranged to be suitable for detection with only statistical  channel knowledge at users as given in \eqref{eqn:k_th_Us_WC_Gaussian_Rx_signal}, where  $\tilde{P_d} \triangleq P_d \left(1-\theta\right)$.
According to  \cite{Ngo2017n,Marzetta2016_Book}, the effective noise can be treated as the worst-case independently distributed Gaussian noise. Then, the signal-to-interference-plus-noise ratio (SINR) at the $k$th user can be written as  \eqref{eqn:SINR_kth_US}. 
 By computing the expectation and variance terms, this SINR  \eqref{eqn:SINR_kth_US} can be derived  as (see Appendix \ref{app:Appendix3})
\addtocounter{equation}{2}
	\begin{eqnarray}\label{eqn:SINR_kth_US_analysis}
	\!\!\!\! \gamma_k = \frac{\tilde{P_d} \left(\sum_{m=1}^{M} \eta_{mk}^{1/2} \rho_{mk}\right)^2}{\tilde{P_d} \sum_{m=1}^{M} \sum_{i=1}^{K}\eta_{mi} \rho_{mi} \zeta_{mk} + 1 }.
	\end{eqnarray}
\noindent Then, an achievable rate of the $k$th user can be defined as
\begin{eqnarray}\label{eqn:kth_US_rate}
	R_k = \frac{\left(1-\alpha\right) \tau_d}{\tau_c} \log[2] {1+ \gamma_k},
\end{eqnarray} 
  where the pre-log factor $\left(1-\alpha\right) \tau_d / \tau_c $ captures the effective potion of the coherence interval for the DL data.

\section{Mitigating   near-far effects  }\label{sec:Max_Min_Fairness_Energy_Rate_Trade_off}

Max-min   power control   has been shown to be optimal in the sense of user-fairness under near-far effects in multi-user  systems \cite{Marzetta2016_Book,Marbach2003, Radunovic2007, Zheng2018}.   In SWIPT, user-fairness must be guaranteed not only in terms of the   rates but also with respect to the harvested energy.   In this section, a   max-min fairness-based energy-rate trade-off analysis is presented  to mitigate the deleterious  near-far effects. 

\subsection{Transmit power control for the PS protocol}\label{sec:Transmit_power_control_for_PS_protocol}

In PS protocol, the data transmission and energy harvesting take place within the same time-slot, and thus, the transmit power must be controlled jointly to maximize the  harvested energy and rate of the weakest user. This requires a  multi-objective optimization problem (MOOP) formulation \cite{Bjornson2014}.  Thus, this MOOP can be formulated as
\begin{subequations} \label{eqn:PS_optim}
	\begin{eqnarray}
	\underset{\eta_{mk} \forall{m,k}}{\text{maximize}} &&\quad \left(\bar{R}_{c,PS}\right)^{w_r} \left(\bar{E}_{c,PS}\right)^{w_e} = \lambda_c \label{eqn:PS_power_control}\\
	\text{subject to}  
	&&C_1: R_k^{w_r} \bar{E}_k^{w_e} \geq \lambda_c,\label{eqn:PS_constraint}\\
	&&C_2: \sum\nolimits_{k=1}^{K} \eta_{mk} \rho_{mk} \leq 1,\label{eqn:Tx_signal_power_constraint}\\
	&&C_3: 0 \leq \eta_{mk}, \label{eqn:power_control_constraint}
	\end{eqnarray}	 
\end{subequations}
 where $\bar{R}_{c,PS}$ and $\bar{E}_{c,PS}$ are the system-wide common target  DL  rate and   the harvested energy, respectively. Moreover, $w_r$ and $w_e$ are the priorities assigned to $\bar{R}_{c,PS}$ and $\bar{E}_{c,PS}$, respectively.   Due to the non-convex nature of the multi-objective function in \eqref{eqn:PS_constraint} \cite{Peressini1988}, an optimal closed-form solution to this MOOP becomes mathematically intractable for a finite number of APs\footnote{This MOOP is solved  in the large AP regime as shown in Section \ref{sec:Asymptotic_Transmit_Power_control_for_PS_protocol}.}. To circumvent this difficulty, a two-step approach is proposed to obtain the max-min optimal energy-rate trade-off for PS protocol. In the first step, by  keeping the PS factor fixed, a common  harvested energy is obtained by maximizing the minimum harvested energy.  In the second step, a common  DL user rate can be derived by maximizing the minimum   user rate for a fixed PS factor. The common rate and harvested energy   are functions of the PS factor. By   solving for the PS factor from the  common rate and then by substituting it to the common harvested energy, a   max-min fairness-based energy-rate trade-off can be derived. 

\subsubsection{Maximizing   minimum harvested energy for PS protocol}

A max-min   power control policy  is formulated  to maximize the minimum harvested energy   for a fixed PS factor   as 
\begin{subequations}\label{eq:PS_HE_mx_min}
	\begin{eqnarray}
	\underset{\eta_{mk} \forall{m,k}}{\text{maximize}}&& \quad \bar{E}_{c,PS} \label{eqn:PS_ power_control_energy}\\
	\text{subject to}  &&
	C_1: \bar{E}_k \geq  \bar{E}_{c,PS} , \label{eqn:energy_constraint}\\
	&&C_2: \sum\nolimits_{k=1}^{K} \eta_{mk} \rho_{mk} \leq 1,\label{eqn:Tx_signal_power_constraint_e}\\
	&&C_3: 0 \leq \eta_{mk}, \label{eqn:power_control_constraint_e}
	\end{eqnarray}	 
\end{subequations}
 where $\bar{E}_{c,PS}$ is the common target for the harvested energy and  $\bar{E}_k$ is given in \eqref{eqn:avg_HE_up_bound} with $\alpha = 0$ for the PS protocol.
As per (\ref{eqn:avg_HE_up_bound}),  $\bar{E}_k$ is a non-decreasing  function of the average received power, and thus,  $\bar{E}_k$   can be replaced by $\E{P_k}$, which is given in \eqref{eqn:avg_power_kth_Us}. Next, the max-min optimization problem in (\ref{eq:PS_HE_mx_min})  can be equivalently reformulated by defining a slack variable $\beta_{mk} \triangleq {\eta_{mk}^{1/2}}$ as   
	\begin{subequations} \label{eqn:PS_power_control_HE}
		\begin{eqnarray}
		\underset{\beta_{mk} \forall{m,k}}{\text{maximize}} &&\quad \bar{P}_{c,PS} \label{eqn:PS_power_control_energy_1}\\
		\text{subject to}  
		&&C_1: \E{P_k} \geq  \bar{P}_{c,PS} , \label{eqn:energy_constraint_1}\\
		&&C_2: \sum\nolimits_{k=1}^{K} \beta_{mk}^2 \rho_{mk} \leq 1,\label{eqn:Tx_signal_power_constraint_e_1}\\
		&&C_3: 0 \leq \beta_{mk}, \label{eqn:power_control_constraint_e_1}
		\end{eqnarray}	 
	\end{subequations}
 where $\bar{P}_{c,PS}$ is the common average received power. The constraints \eqref{eqn:energy_constraint_1}-\eqref{eqn:power_control_constraint_e_1} and the objective function \eqref{eqn:PS_power_control_energy_1} are convex \cite{Boyd2004}.
 This convex optimization problem (\ref{eqn:PS_power_control_HE}) can be modeled as a second order cone programming (SOCP) and solved via CVX \cite{Boyd2004}.

\begin{algorithm}\label{algorithm1}
	\caption{The proposed statistical channel information based pilot assignment and user grouping algorithm}
	\begin{algorithmic}[1]
		\renewcommand{\algorithmicrequire}{\textbf{Input:}}
		\renewcommand{\algorithmicensure}{\textbf{Output:}}
		\REQUIRE Path-losses between all  $M$ APs and  $K$ users and the average transmit power, $P_d$.
		\ENSURE  The power control coefficients $\eta_{mk}$ for $m \in \{1, \cdots, M\} $ and $k \in \{1, \cdots, K\}$. 
		\STATE Begin - CVX.
		\STATE  Set a common target average received power, $\bar{P}_{c,PS}$.
		\STATE Solve the convex feasibility problem  given by
		\begin{eqnarray} 
		&&\| \boldsymbol{V}_{E_k} \|  \geq  \sqrt{{\bar{P}_{c,PS}}/{P_d}},
		\end{eqnarray}
		\noindent
		which is subjected to the constraints \eqref{eqn:Tx_signal_power_constraint_e_1}-\eqref{eqn:power_control_constraint_e_1}.
		Moreover, $\| \boldsymbol{V}_{E_k} \|  \triangleq \left[ \mathbf{v}_{E1}^T,  \; \mathbf{v}_{E2}^T\right] $, where $\mathbf{v}_{E1}$ and  $\mathbf{v}_{E2}$ are given by
		\begin{subequations}\label{eqn:vE1_1}
			\begin{eqnarray}
			\!\!\!\!\!\! \mathbf{v}_{E1} &\triangleq& \left[ \beta_{1k} \rho_{1k} , \cdots , \beta_{Mk} \rho_{Mk} \, \right]^T,  \\
			\!\!\!\!\!\! \mathbf{v}_{E2} &\triangleq&  \left[ \beta_{11} \sqrt{\rho_{11} \zeta_{1k}} , \cdots , \beta_{MK} \sqrt{\rho_{MK} \zeta_{Mk}} \, \right]^T.
			\end{eqnarray}
		\end{subequations}
		\STATE  End - CVX.
		
		\RETURN $\eta_{mk} = \beta_{mk}^2 $ for $m \in \{1, \cdots, M\}$ and $k \in \{1, \cdots, K\}$.
	\end{algorithmic} 
\end{algorithm}

\subsubsection{Maximizing   minimum achievable rate for PS protocol}
 
By invoking the max-min optimization \cite{Marzetta2016_Book},   the optimal transmit power control coefficients can be computed for a fixed PS factor to maximize the minimum   DL rate among all users. To this end, a max-min  power control problem can be formulated by replacing $\bar{E}_{c,PS}$ and $\bar{E}_{k}$ in  \eqref{eq:PS_HE_mx_min} with $\bar{R}_{c,PS}$, the common/target  DL  rate that can be achieved at the optimal solution and $R_k$, the instantaneous DL rate of the $k$th user  given in \eqref{eqn:kth_US_rate} with $\alpha = 0$, respectively. 

Since  $R_k$ is a  non-decreasing  function of its argument,  the $k$th user rate  can be replaced by the  SINR of the $k$th user $(\gamma_k)$, which is given in  \eqref{eqn:SINR_kth_US_analysis}. 
By introducing a common SINR  $(\gamma_{c,PS})$ for all users and by defining a slack variable $\beta_{mk} \triangleq {\eta_{mk}^{1/2}} $, the optimization problem  can be equivalently reformulated as
		\begin{subequations} \label{eqn:PS_power_control_SNR_1}
			\begin{eqnarray}
			\underset{\beta_{mk} \forall{m,k}}{\text{maximize}}	&& \quad \gamma_{c,PS} \label{eqn:PS_power_control_1_rate}\\
			\text{subject to}    
			&&C_1: \gamma_k \geq \gamma_{c,PS}, \label{eqn:SINR_constraint}\\
			&&C_2: \sum\nolimits_{k=1}^{K} \beta_{mk}^2 \rho_{mk} \leq 1,\label{eqn:Tx_signal_power_constraint_1}\\
			&&C_3: 0 \leq \beta_{mk}.\label{eqn:power_control_constraint_1}
			\end{eqnarray}	 
		\end{subequations}

Since the objective function in \eqref{eqn:PS_power_control_1_rate} is a quasi-concave function, the underlying  optimization problem is also quasi-concave \cite{Boyd2004,Ngo2017}. Then, an optimal solution for the feasibility problem in (\ref{eqn:PS_power_control_SNR_1}) for a fixed PS factor ($\theta$) can be obtained by invoking a variation of the Bisection method for solving SOCP \cite{Boyd2004,Ngo2017} as presented in Algorithm 2.

\setcounter{mycounter}{\value{equation}}
\begin{figure*}[!t] 
	\vspace{-7mm}
	\addtocounter{equation}{0}
	\begin{eqnarray}\label{eqn:PS_trade_off}
	R_{c,PS}^* = \frac{ \tau_d}{\tau_c}  \log[2]{1 + \frac{P_d \left(1-\bar{E}_{c,PS}^* \Psi_{EH}^{-1}\left(\bar{P}_{c,PS}^*\right) / \tau_d\right) \left(\sum_{m=1}^{M} \eta_{mk}^{1/2} \rho_{mk}\right)^2}{P_d \left(1-\bar{E}_{c,PS}^* \Psi_{EH}^{-1}\left(\bar{P}_{c,PS}^*\right) / \tau_d\right) \sum_{m=1}^{M} \sum_{i=1}^{K}\eta_{mi} \rho_{mi} \zeta_{mk} + 1 }}
	\end{eqnarray} 
	\vspace{-8mm}
\end{figure*}
\setcounter{equation}{\value{mycounter}}
\setcounter{mycounter}{\value{equation}}
\begin{figure*}[!t] 
	\addtocounter{equation}{3}
	\begin{eqnarray}\label{eqn:TS_trade_off}
	R_{c,TS}^* = \frac{\left(1- \bar{E}_{c,TS}^* \Psi_{EH}^{-1} \left(\bar{P}_{c,TS}^*\right) /\tau_{d}\right) \tau_d}{\tau_c} \log[2] {1+   \frac{P_d  \left(\sum_{m=1}^{M} \eta_{mk}^{1/2} \rho_{mk}\right)^2}{P_d  \sum_{m=1}^{M} \sum_{i=1}^{K}\eta_{mi} \rho_{mi} \zeta_{mk} + 1 }}
	\end{eqnarray}
	\vspace{-2mm}
	\hrulefill
\end{figure*}
\setcounter{equation}{\value{mycounter}}

\subsubsection{Max-min optimal energy-rate trade-off for PS protocol}\label{sec:Max_min_optimal_energy_rate_trade_off_for_PS_protocol}
By first solving for $\theta$ in \eqref{eqn:energy_constraint} and then substituting it into \eqref{eqn:SINR_constraint},    a max-min fairness-based common energy-rate trade-off for PS protocol can be obtained by using the max-min optimal solutions of $\bar{P}_{c,PS}^*$ and $\gamma_{c,PS}^*$  as in \eqref{eqn:PS_trade_off}.
 
\textbf{\textit{Remark 2:}} Any point along this 
  energy-rate trade-off curve  (\ref{eqn:PS_trade_off}) corresponds to a  common harvested energy and user rate pair computed via solving max-min fairness-based optimization. Then, the PS factor ($\theta$) for any point along this   trade-off curve can be   obtained via back-substitution of either common harvested energy or user rate into the corresponding metric, which is in turn a function of the PS factor.

\vspace{-3mm}

\begin{algorithm}\label{algorithm2}
	\caption{: Bisection algorithm for transmit power control}
	\begin{algorithmic}[1]
		\renewcommand{\algorithmicrequire}{\textbf{Input:}}
		\renewcommand{\algorithmicensure}{\textbf{Output:}}
		\REQUIRE Path-losses between all  $M$ APs and  $K$ users and the average transmit power, $P_d$.
		\ENSURE  The power control coefficients $\eta_{mk}$ for $m \in \{1, \cdots, M\} $ and $k \in \{1, \cdots, K\}$. \\
		\textbf{Initialization}: 
		Define an initial region for the objective functions by choosing appropriate values for $t_{min}$ and $t_{max}$. Choose a tolerance $\epsilon > 0$.

		\WHILE{$ t_{max} - t_{min}  > \epsilon$}
		\STATE Calculate, $\gamma_{c,PS} = {\left(t_{max}+t_{min}\right)}/{2}$. 
		\STATE Solve the convex feasibility problem, which can be given as
\addtocounter{equation}{1}
			\begin{eqnarray} 
			&&\| \boldsymbol{V}_{\gamma_k} \| \leq \frac{1}{\sqrt{\gamma_{c,PS}}} \left(\sum\nolimits_{m=1}^{M} \beta_{mk} \rho_{mk}  \right),
			\end{eqnarray}
		which is subjected to $C_2$ and $C_3$ given in \eqref{eqn:Tx_signal_power_constraint_1} and \eqref{eqn:power_control_constraint_1}, respectively. Moreover, $\boldsymbol{V}_{\gamma_k} \triangleq  \left[\mathbf{v}_{1}^T,  \; {1}/{\sqrt{\tilde{P_d}}} \right]$, where 
			\begin{eqnarray}\label{eqn:v1}
			\mathbf{v}_{1} &\triangleq& \left[ \beta_{11} \sqrt{\rho_{11}} , \cdots , \beta_{MK} \sqrt{\rho_{MK}} \, \right]^T.
			\end{eqnarray} 
		
		\STATE If the status of the problem is feasible, then set $t_{min} = \gamma_{c,PS}$, otherwise set $t_{max} = \gamma_{c,PS}$.
		
		\STATE Stop if $t_{max}-t_{min} < \epsilon$. Otherwise go to Step 2.
		\ENDWHILE
		\RETURN $\eta_{mk} = \beta_{mk}^2 $ for $m \in \{1, \cdots, M\}$ and $k \in \{1, \cdots, K\}$.
	\end{algorithmic} 

\vspace{-0mm}
	\end{algorithm}

\subsection{Transmit power control  for the TS protocol}\label{sec:Transmit_power_control_of_energy_harvesting_for_TS_protocol}

In this subsection, our objective is to mitigate the adverse near-far effects   of SWIPT with TS protocol, in which  the power and information transfers occur in two orthogonal time-slots. Thus, the transmit power can be optimized to ensure max-min user-fairness in terms of both harvested energy and user rates for a given TS factor. Thereby, the max-min fairness optimal energy-rate trade-off can be derived.   By traversing through this trade-off curve, a max-min optimal TS factor $(\alpha)$ for any  common energy-rate can be obtained.

\subsubsection{Max-min fairness optimal common harvested energy for TS protocol}

In  this subsection,   a  common harvested energy at each user    is obtained via  max-min  transmit power control. 
Since all users achieve this common harvested energy for a given TS factor,  the  near-far effect can be readily mitigated. 
Since the harvested energy  is a non-decreasing function  of the average received power, the max-min optimal  power control coefficients can be computed  by setting a common received energy $(\bar{E}_{c,TS})$ for every user. Thus, a max-min power control policy to ensure a common harvested energy can be formulated by replacing the $\bar{E}_{c,PS}$ and $\bar{E}_{k}$ in  \eqref{eq:PS_HE_mx_min} with $\bar{E}_{c,TS}$ and $ \E{P_k}$, defined in (\ref{eqn:avg_power_kth_Us}), respectively.
The objective function of the optimization problem can shown to be  convex \cite{Marzetta2016_Book,Boyd2004}. The equality of the respective optimization problem holds at the optimal solution, and thus, all harvested energies must be equal to the common maximum value, $\bar{E}_{c,TS}^*$. Then, the optimization problem  can be reformulated and then solved by adopting an algorithm  similar to one used for \eqref{eqn:PS_power_control_HE} by using CVX \cite{Boyd2004}.

\subsubsection{Max-min fairness optimal common user rate for TS protocol}\label{sec:Transmit_power_control_of_rate_for_TS_protocol}

A max-min fairness common   rate for a fixed  TS factor ($\alpha$) can be obtained by computing transmit 
power allocation coefficients  to maximize the minimum   DL user rates among all users \cite{Marzetta2016_Book}. The user rate in  \eqref{eqn:kth_US_rate} is a non-decreasing  function of the  SINR, $\gamma_k$. Thus, by introducing a common SINR $(\gamma_{c,TS})$ for all  users,  the power control coefficients can be computed  such that this common SINR is maximized. Then, by keeping the TS factor fixed,  a max-min power control policy can be  formulated by replacing $\gamma_{c,PS}$ in \eqref{eqn:PS_power_control_SNR_1} with $\gamma_{c,TS}$.
 Since the objective function  is quasi-concave, this optimization problem too becomes quasi-concave \cite{Boyd2004,Ngo2017}. Then, this can be solved via the same Bisection method \cite{Boyd2004} proposed for (\ref{eqn:PS_power_control_SNR_1}).

\subsubsection{Max-min optimal energy-rate trade-off for TS protocol}\label{sec:Max_min_optimal_energy_rate_trade_off_for_TS_protocol}

The max-min optimal energy-rate trade-off for the TS protocol  can be derived by first solving for $\alpha$ in \eqref{eqn:avg_HE_up_bound} and then by substitute it into \eqref{eqn:kth_US_rate}. Thus, the   max-min optimal  energy-rate trade-off for the TS protocol can be written as   \eqref{eqn:TS_trade_off}.

\textbf{\textit{Remark 3:}} By traversing through the energy-rate trade-off curve in (\ref{eqn:TS_trade_off}), the max-min optimal TS factor for any pair of common energy and rate can be obtained.  The TS factor ($\alpha$) is a function of the energy harvesting duration within a  coherence interval. Thus, our solution inherently deals with optimizing the time-slots allocated for energy harvesting and data transmission in the sense of  max-min user-fairness.

\setcounter{mycounter}{\value{equation}}
\begin{figure*}[!t] 
	\vspace{-7mm}
	\addtocounter{equation}{6}
	\begin{eqnarray}\label{eqn:asymptotic_PS_trade_off}
R_{c,PS}^{\infty*} = \frac{\tau_d}{\tau_c} \log[2] {1 + \left(1 - E_{c,PS}^{\infty *} \Psi_{EH}^{-1} \left(P_{c,PS}^{\infty *}\right)/\tau_d\right) P_d \left(\sum_{m=1}^{M} \eta_{mk}^{1/2} \rho_{mk} \right)^2} 
\end{eqnarray}
	\vspace{-9mm}
\end{figure*}
\setcounter{equation}{\value{mycounter}}
\setcounter{mycounter}{\value{equation}}
\begin{figure*}[!t] 
	\addtocounter{equation}{7}
\begin{eqnarray}\label{eqn:asymptotic_TS_trade_off}
R_{c,TS}^{\infty*} = \frac{\left(1 - E_{c,TS}^{\infty *}\Psi_{EH}^{-1} \left(P_{c,TS}^{\infty *}\right)/ \tau_d \right)\tau_d}{\tau_c} \log[2] {1 +  P_d \left(\sum_{m=1}^{M} \eta_{mk}^{1/2} \rho_{mk} \right)^2}
\end{eqnarray}
	\vspace{-2mm}
	\hrulefill
\end{figure*}
\setcounter{equation}{\value{mycounter}}

\section{Asymptotic Analysis for DL transmission}\label{sec:Asymptotic_Analysis}

In this section, the performance of SWIPT  is investigated when the number of APs   grows without bound ($M\rightarrow \infty$). 
By substituting the UL channel estimate in  \eqref{eqn:pilot_estimate} and \eqref{eqn:estimate_of_h_mk} into  \eqref{eqn:rearranged_kth_Rx_signal}, an asymptotic 
approximation  for the received power at the $k$th user   can be derived as (see Appendix \ref{app:Appendix5})
\addtocounter{equation}{1}
\begin{eqnarray}\label{eqn:asymptotic_Rx_power}
	P_{k,{\infty}} = P_d \left(\sum_{m=1}^{M} \eta_{mk}^{1/2} \rho_{mk} \right)^2.
\end{eqnarray}
 The asymptotic harvested energy at the $k$th user   is given by
\begin{eqnarray}\label{eqn:asymptotic_HE}
\!\!\!\!\!	\!\!\!\!\!\bar{E}_k \!=\!  \alpha \tau_d \Psi_{EH} \left( \E{P_{k,{\infty}}} \right) + (1-\alpha) \tau_d \Psi_{EH} \left(\theta \E{ P_{k,{\infty}}}\right).
\end{eqnarray} 
 The asymptotic achievable rate at the $k$th user is given by 
\begin{eqnarray}\label{eqn:asymptotic_rate}
	R_{k,\infty} = \frac{\left(1-\alpha\right) \tau_d}{\tau_c} \log[2] {1 + \gamma_{k,\infty}} ,
\end{eqnarray} 
 where the asymptotic SINR ($\gamma_{k,\infty}$) is given by
\begin{eqnarray}\label{eqn:asymptotic_SNR}
	\gamma_{k,\infty} = \tilde{P_d} \left(\sum_{m=1}^{M} \eta_{mk}^{1/2} \rho_{mk} \right)^2,
\end{eqnarray}
 where the receiver noise variance  is normalized to unity.

\subsection{Asymptotic transmit power control  for PS protocol}\label{sec:Asymptotic_Transmit_Power_control_for_PS_protocol}

In contrast to  Section \ref{sec:Transmit_power_control_for_PS_protocol}, in this subsection,   a MOOP \cite{Bjornson2014} is formulated to control transmit power in the asymptotic AP regime ($M\rightarrow \infty$). This MOOP formulation is particularly useful for the PS protocol in which the energy and  data transfer  take place in the same time slot. Thus, a MOOP is formulated to asymptotically optimize the energy-rate trade-off of   PS protocol as   \cite{Bjornson2014}
\begin{subequations} \label{eqn:asymptotic_PS_optim}
	\begin{eqnarray}
	\underset{\beta_{mk} \forall{m,k}}{\text{maximize}} &&\quad \left({P}_{c,PS}^{\infty}\right)^{w_e} \left(\gamma_{c,PS}^{\infty}\right)^{w_r} = \lambda \label{eqn:asymptotic_PS_power_control}\\
	 \text{subject to}  
	&&C_1: \sqrt{\lambda} \leq\! \sqrt{P_d}\! \left(\sum\nolimits_{m=1}^{M} \beta_{mk} \rho_{mk}\!\right),\label{eqn:asymptotic_PS_constraint}\\
	&&C_2: \sum\nolimits_{k=1}^{K} \beta_{mk}^2 \rho_{mk} \leq 1,\label{eqn:asymptotic_Tx_signal_power_constraint}  \\
	&&C_3: 0 \leq \beta_{mk}, \label{eqn:asymptotic_power_control_constraint}
	\end{eqnarray}	 
\end{subequations}
 where $w_e$ and $w_r$ are the priorities assigned for the harvested energy and rate. For instance, if  energy and data transfers   have the same priority, then  $w_e=w_r=1$ \cite{Bjornson2014,Kudathanthirige2019}. In (\ref{eqn:asymptotic_PS_constraint}),   $\beta_{mk} \triangleq {\eta_{mk}^{1/2}}$ and $\lambda$ is a slack variable for the common rate and harvested energy. All constraints and   objective function  in  \eqref{eqn:asymptotic_PS_optim} are convex \cite{Marzetta2016_Book,Boyd2004}. Consequently, the MOOP in \eqref{eqn:asymptotic_PS_optim} is solved via CVX \cite{Boyd2004}.

At the optimal solution,  by solving for $\theta$ in the harvested energy in \eqref{eqn:asymptotic_HE} and then substituting it into the  user rate in \eqref{eqn:asymptotic_SNR}, the asymptotically optimal energy-rate trade-off is derived as   \eqref{eqn:asymptotic_PS_trade_off}.

\subsection{Asymptotic transmit power control  for TS protocol}\label{sec:Asymptotic_Transmit_Power_control_for_TS_protocol}

Since the energy  and data transmissions take place in two orthogonal time-slots in the TS protocol, the  
two separate max-min problems to optimize the harvested energy and achievable rate can be formulated by following steps similar to (\ref{eqn:asymptotic_PS_optim}). 
For the energy transfer phase of  the TS protocol, $\lambda$ in the objective function  in  (\ref{eqn:asymptotic_PS_optim}) becomes the common harvested energy with unity priority; $w_e=1$ ($w_r=0$). 
For the data transfer phase, $\lambda$ in the objective function  in  (\ref{eqn:asymptotic_PS_optim}) must be replaced by  the common achievable rate  with unity priority; $w_r=1$ ($w_e=0$). These optimization problems can be shown to be convex and hence can be solved via CVX \cite{Boyd2004}.

By following a similar approach to the PS protocol, at the optimal solution,  we derive the asymptotic max-min optimal energy-rate trade-off as $M\rightarrow \infty$ as  in \eqref{eqn:asymptotic_TS_trade_off}.

\section{Uplink Transmission}\label{sec:UL_tx}


\subsection{Signal model for UL transmission}\label{sec:Signal_model_UL}

In UL,   the users are typically assumed to be  equipped with an external back-up power source in the event that the harvested energy level  is  inadequate for  sending UL data in a mission critical application \cite{Ng2013,Yao2017}.     Thus,  the total energy at the $k$th user can be written as
\addtocounter{equation}{2}
\begin{eqnarray}\label{eqn:E_k_total}
 E_{k}^{Tot} = E_{k}^{Rem} + \bar{E}_k, 
 \end{eqnarray}
  where $E_{k}^{Rem}$ is the back-up/remaining energy at the $k$th user and $\bar{E}_k$ is the total harvested energy    \eqref{eqn:average_HE}. We  also defined a threshold energy, $E_{k}^{Min}$, which is the minimum energy that is  needed for pilot transmission at the $k$th user. Then, the $k$th user transmits its UL signal if 
 \begin{eqnarray}\label{eqn:E_k_tx_constraint}
 E_{k}^{Min} > (1 - \kappa)E_{k}^{Tot} , 
 \end{eqnarray}
  where $0 < \kappa < 1$ is a fraction used to allocate     transmit power between the UL payload data and  pilots. 
Then, a predefined  portion of this total   energy  can be used for   UL payload data transfer at each user, while the remaining portion can be used for UL pilot transfer.
The average transmit power used for UL data transfer at the $k$th user can be derived as
\begin{eqnarray}\label{eqn:p_uk}
	P_{uk} = {\kappa E_{k}^{Tot} }/{\tau_{u}}, 
\end{eqnarray} 
where  $\tau_u$ is the coherence interval allocated for UL data transmission, and  $E_{k}^{Tot}$ is the total available energy defined in \eqref{eqn:E_k_total}. 
The signal received at the $m$th AP can be written as
\begin{eqnarray}\label{eqn:Rx_m_AP}
	{r}_{m,u} =  \sum_{k=1}^{K} \sqrt{P_{uk} \eta_{k}} h_{mk} q_{k,u} + n_{m,u},
\end{eqnarray} 
 where $q_{k,u}$ is the signal transmitted by the $k$th user   satisfying $\E{|q_{k,u}|^2} = 1$, $\eta_k$ is its UL power control coefficient and $n_{m,u} \sim \mathcal{CN}(0,1)$ is an AWGN at the $m$th AP.  
Then, the $m$th AP employs a conjugate precoder, which is constructed based on its local  UL channel estimate \eqref{eqn:estimate_of_h_mk}. The post-processed signal   at the CPU  from the $k$th user  can be written  as
		\begin{eqnarray}\label{eqn:Rx_CPU}
		{r}_{k,u} =  \sum_{m=1}^{M} \hat{h}_{mk}^* {r}_{m,u},
		\end{eqnarray}
		  where ${r}_{m,u}$ is given in \eqref{eqn:Rx_m_AP}. 
	By substituting (\ref{eqn:Rx_m_AP}) into (\ref{eqn:Rx_CPU}), the  received signal at the CPU can be rewritten   as 
		\begin{eqnarray}\label{eqn:Rx_m_CPU_rearrenge}
		{r}_{k,u} &=&  \sum_{m=1}^{M}  \sqrt{P_{uk} \eta_{k}} \hat{h}_{mk}^* h_{mk} q_{k,u} +  \sum_{m=1}^{M} \hat{h}_{mk}^* n_{m,u} \nonumber \\
		&&+ \sum_{m=1}^{M} \sum_{i \neq 1}^{K}  \sqrt{P_{ui} \eta_{i}} \hat{h}_{mk}^* h_{mi} q_{i,u} ,
		\end{eqnarray}
 where the first, second, and third terms represent the desired signal component, filtered noise and, inter-user interference.

\subsection{The UL achievable rate}\label{sec:Achievable_UL_rate}

By following steps similar to (\ref{eqn:k_th_Us_WC_Gaussian_Rx_signal}), the   received signal \eqref{eqn:Rx_m_CPU_rearrenge} can be rearranged to make it suitable for decoding at the CPU as   \eqref{eqn:k_CPU_WC_Gaussian_Rx_signal}.
 By using  \eqref{eqn:k_CPU_WC_Gaussian_Rx_signal}, the effective UL SINR can be derived as   \eqref{eqn:SINR_CPU}.
The variance and expectation terms of (\ref{eqn:SINR_CPU}) can be computed by using techniques similar to those in Appendix \ref{app:Appendix3}. Thereby, the UL SINR can be derived in closed-form as 
\addtocounter{equation}{2}
\begin{eqnarray}\label{eqn:SINR_CPU_analysis}
	\!\!\!\! \gamma_{k,u} = \frac{P_{uk} \eta_{k} \left(\sum_{m=1}^{M}  \rho_{mk}\right)^2}{  \sum_{i=1}^{K} P_{ui} \eta_{i} \sum_{m=1}^{M}  \rho_{mk} \zeta_{mi} + \sum_{m=1}^{M} \rho_{mk} }.
\end{eqnarray}
Next, we derive the  achievable UL rate  as follows:
\begin{eqnarray}\label{eqn:UL_rate}
	R_{k,u} = \frac{\tau_u}{\tau_c} \log[2] {1+ \gamma_{k,u}},
\end{eqnarray} 
where $\tau_u/\tau_c$ captures the effective potion of the coherence interval used for UL data transfer and $\gamma_{k,u}$ is defined in \eqref{eqn:SINR_CPU_analysis}.

\subsection{Joint DL/UL transmit power control}\label{sec:Transmit_power_control_for_UL_transmission}

To ensure that the UL  user rates  are independent of the near-far effects, while maximizing the minimum  DL harvested energy, a joint max-min fairness optimal DL/UL  transmit power control problem is formulated. Thus, at the  optimal solution of this joint power control, all users achieve a common UL rate regardless of their channel conditions, while satisfying  a common  DL harvested energy constraint for all users.   The achievable rate  in \eqref{eqn:UL_rate} is   a monotonically increasing function of its argument $\gamma_{k,u}$. 
Since the non-convexity of   non-linear energy harvesting model in   \eqref{eqn:EH_model}, we adopted linear energy harvesting model only for calculating UL transmit powers in the joint optimization problem. By first introducing a common UL SINR ($\gamma_{c,u}$) and a common DL harvested energy ($E_{c,d}$) for all users and then by defining  slack variables $\beta_k \triangleq {\eta_k^{1/2}}$ and $\beta_{mk} \triangleq \eta_{mk}$, 
a max-min joint   power control problem can be formulated as 
\begin{subequations} \label{eqn:power_control_UL_SNR}
	\begin{eqnarray}
	\!\!\!\!\!\!\!\!\!\!\!\!\underset{\beta_{mk}, \beta_{k}  \forall{m,k}}{\text{maximize}}&& \quad (E_{c,d})^{w_e}  (\gamma_{c,u})^{w_r}  \label{eqn:UL_power_control_rate}\\
	\text{subject to} \nonumber \\
	&&\!\!\!\!\!\! \!\!\!\!\!\! \!\!\!\!\!\! \!\!\!\!\!\!C_1: \E{P_k} \geq  E_{c,d},  \label{eqn:DL_energy}\\
	&&\!\!\!\!\!\! \!\!\!\!\!\! \!\!\!\!\!\! \!\!\!\!\!\!C_2: \| \boldsymbol{V}_{\gamma_{k,u}} \| \leq \frac{1}{\sqrt{\gamma_{c,u}}} \left(\sum\nolimits_{m=1}^{M} \beta_{k} \rho_{mk}  \right), \label{eqn:UL_SINR_constraint}\\
	&&\!\!\!\!\!\! \!\!\!\!\!\! \!\!\!\!\!\! \!\!\!\!\!\!C_3: \sum\nolimits_{k=1}^{K} \beta_{mk} \rho_{mk} \leq 1, \quad \text{and} \quad  0 \leq \beta_{mk} \leq 1,\label{eqn:eng_constraint}\\
	&&\!\!\!\!\!\! \!\!\!\!\!\! \!\!\!\!\!\! \!\!\!\!\!\!C_4: \sum\nolimits_{k=1}^{K} \beta_{k}^2 \rho_{mk} \leq 1, \quad \text{and} \quad 0 \leq \beta_{k} \leq 1, \label{eqn:UL_Tx_signal_power_constraint_1}
	\end{eqnarray}	 
\end{subequations}
where $\boldsymbol{V}_{\gamma_{k,u}} \triangleq \left[\mathbf{v}_{1,u}^T,  \; \mathbf{v}_{2,u}^T \right]$. Moreover,  $\mathbf{v}_{1,u}$ and $\mathbf{v}_{2,u}$ are given  by
\begin{subequations}
	\begin{eqnarray}
	\!\!\!\!\!\! \!\!\! \mathbf{v}_{1,u} &\triangleq& \left[ \beta_{1} \sqrt{\frac{P_{u1}  {\rho_{1k} \zeta_{11}}}{P_{uk}}}   , \cdots , \beta_{K} \sqrt{\frac{P_{uK}  {\rho_{Mk} \zeta_{MK}}}{P_{uk}}}  \right]^T \!\!, \\
	 	\!\!\!\!\!\! \!\!\! \mathbf{v}_{2,u} &\triangleq& \left[ \sqrt{\rho_{1k} / P_{uk}} , \cdots , \sqrt{\rho_{Mk} / P_{uk}} \, \right]^T. \label{eqn:v_UL_1} 
	\end{eqnarray}
\end{subequations}
Since the objective function in \eqref{eqn:UL_power_control_rate} is quasi-concave, this optimization problem is also quasi-concave \cite{Boyd2004,Ngo2017}. Thus,  the optimization problem in \eqref{eqn:power_control_UL_SNR} can be solved as a CVX \cite{Boyd2004} feasible problem by using Algorithm 3. Since both energy and rate are equally prioritized in SWIPT,  we set the weights in (\ref{eqn:UL_power_control_rate}) to $w_e=w_r=1$ \cite{Bjornson2014}.

\begin{algorithm}\label{algorithm3}
	\caption{: Bisection algorithm for   DL/UL   power control}
	\begin{algorithmic}[1]
		\renewcommand{\algorithmicrequire}{\textbf{Input:}}
		\renewcommand{\algorithmicensure}{\textbf{Output:}}
		\REQUIRE Path-loss coefficients between all  $M$ APs and  $K$ users,  the average transmit powers of the users, and UL energy conversion efficiencies of the users.
		\ENSURE  The DL power control coefficients $\eta_{mk}$ for $m \in \{1, \cdots, M\} $, $k \in \{1, \cdots, K\}$ and the UL power control coefficients $\eta_{k}$ for $k \in \{1, \cdots, K\}$.  \\
		\textbf{Stage 1}: 
		
		\STATE Begin - CVX.
		\STATE Set a common target value for harvested energy.
		\STATE Solve the convex feasible problem given by
		\begin{eqnarray} 
		&&\| \boldsymbol{V}_{E_k} \|  \geq  \sqrt{{ E_{c,d}}/{P_d}},
		\end{eqnarray}
 which is subjected to the constraint \eqref{eqn:eng_constraint}.
		 $\| \boldsymbol{V}_{E_k} \|  \triangleq \left[ \mathbf{v}_{E1}^T,  \; \mathbf{v}_{E2}^T\right] $, where $\mathbf{v}_{E1}$ and  $\mathbf{v}_{E2}$ are given in \eqref{eqn:vE1_1}. 
		
		\STATE Compute the UL transmit power of the user via DL power control coefficients ($\beta_{mk}^*$) by assuming that only harvested energy is used for UL data transmission
		\begin{eqnarray} 
		\!\!\!\!\!\!\! P_{uk}^* =  \Omega  \left(   \sum_{m=1}^{M}  \beta_{mk}^{*} \rho_{mk}^2   +    \sum_{m=1}^{M}  \sum_{i=1}^{K}  \beta_{mi}^* \rho_{mi} \zeta_{mk}   \right),
		\end{eqnarray}
		where $\Omega \triangleq  {\kappa \tau_{d}  \left(\alpha + \left(1 - \alpha\right)  \theta\right) P_d }/{\tau_{u}}$.

		\textbf{Stage 2}:
		
		\STATE By using Algorithm 2, solve the the convex feasible problem given by
		\begin{eqnarray}
		\| \boldsymbol{V}_{\gamma_{k,u}} \| \leq \frac{1}{\sqrt{\gamma_{c,u}}} \left(\sum_{m=1}^{M} \beta_{k} \rho_{mk}  \right), 
		\end{eqnarray}
 which is subjected to \eqref{eqn:UL_Tx_signal_power_constraint_1}. Further, 
		  $\boldsymbol{V}_{\gamma_{k,u}} \triangleq \left[\mathbf{v}_{1,u}^T,  \; \mathbf{v}_{2,u}^T \right]$, where $\mathbf{v}_{1,u}$ and $\mathbf{v}_{2,u}$ are given  in \eqref{eqn:v_UL_1}. 
		
		\STATE End - CVX.
		
		\RETURN $\eta_{mk} = \beta_{mk} $ for $m \in \{1, \cdots, M\}$ and $k \in \{1, \cdots, K\}$ and $\eta_{k} = \beta_{k}^2$ for $k \in \{1, \cdots, K\}$.
	\end{algorithmic} 
\end{algorithm}

\setcounter{mycounter}{\value{equation}}
\begin{figure*}[!t] 
	\vspace{-7mm}
	\addtocounter{equation}{-9}
	\begin{eqnarray}\label{eqn:k_CPU_WC_Gaussian_Rx_signal}
	r_{k,u} &=& \underbrace{ \E {\sum_{m=1}^{M} \sqrt{P_{uk} \eta_{k}} \hat{h}_{mk}^* {h}_{mk} q_{k,u}}  }_{\text{desired signal}} + \underbrace{  \left(\sum_{m=1}^{M} \sqrt{P_{uk} \eta_{k}} \hat{h}_{mk}^* {h}_{mk} q_{k,u} - \E{\sum_{m=1}^{M} \sqrt{P_{uk} \eta_{k}} \hat{h}_{mk}^* {h}_{mk} q_{k,u} } \right) }_{\text{detection uncertainty}} \nonumber \\
	&&+ \underbrace{ \sum_{m=1}^{M} \sum_{i \neq k}^{K} \sqrt{P_{ui} \eta_{i}} \hat{h}_{mk}^* {h}_{mi} q_{i,u} }_{\text{Inter-user interference}} + \underbrace{\sum_{m=1}^{M} \hat{h}_{mk}^* n_{m,u}}_{\text{AWGN}}
	\end{eqnarray}
	\vspace{-9mm}
\end{figure*}
\setcounter{equation}{\value{mycounter}}
\setcounter{mycounter}{\value{equation}}
\begin{figure*}[!t] 
	\addtocounter{equation}{-8}
	\begin{eqnarray}\label{eqn:SINR_CPU}
	\!\!\!\!\! \!\!\!\!\!
	\gamma_{k,u} =  \frac{  \left| \E{\sum_{m=1}^{M} \sqrt{P_{uk} \eta_{k}}  \hat{h}_{mk}^* h_{mk} } \right|^2}{  \Var{\sum_{m=1}^{M} \sqrt{P_{uk} \eta_{k}}  \hat{h}_{mk}^* h_{mk} } + \E{\left|\sum_{m=1}^{M} \sum_{i \neq k}^{K} \sqrt{P_{ui} \eta_{i}}  \hat{h}_{mk}^* h_{mi} \right|^2 } + \E{\left|\sum_{m=1}^{M} \hat{h}_{mk}^* n_{m,u}\right|^2} }
	\end{eqnarray}
	\vspace{-0mm}
	\hrulefill
\end{figure*}
\setcounter{equation}{\value{mycounter}}

\subsection{Computational complexity}\label{sec:compt_cmpx}

 The convex optimization problems   in \eqref{eqn:PS_power_control_HE}, \eqref{eqn:PS_power_control_SNR_1}, \eqref{eqn:asymptotic_PS_optim}, and \eqref{eqn:power_control_UL_SNR} are typically solved as SOCP via Matlab CVX \cite{Boyd2004}. Thus, the computational complexity in terms of the number of arithmetic operations   to solve the corresponding SOCP can be given as $\mathcal{O} (n_v^2 n_c)$, where $n_v$ is the number of optimization variables and $n_c$ is the number of second order cone constraints \cite{Lobo1998}. The total number of iteration required for solving the    Bisection method is given by $\log[2]{{[t_{max}-t_{min}]}/{2}}$ \cite{Boyd2004}, where $t_{max}$ and $t_{min}$ are defined in Algorithm 2.  Thus, these computational complexities are furnished   in Table  \ref{tab:complexity}. 

\begin{table} [!h]
	\caption{Computational complexity of the proposed algorithms in terms of the number of arithmetic operations }
	\vspace{-6mm}
	\begin{center}\label{tab:complexity}
		\begin{tabular}{|c|c|}
			\hline
			Algorithm & Number of arithmetic operations  \\
			\hline
			Algorithm 1 &  $\mathcal{O} (M^2 K^3)$ \\
			\hline
			Algorithm 2 &  $\log[2]{{[t_{max}-t_{min}]}/{2}} \times \mathcal{O} (M^2 K^3)$ \\
			\hline
			Algorithm 3 &  $\log[2]{{[t_{max}-t_{min}]}/{2}} \times \mathcal{O} (M^2 K^3)^2$ \\
			\hline
		\end{tabular}
	\end{center} 
\vspace{-6mm}
\end{table}

\section{The implications  of DL pilot transmission and DL channel estimates at users}\label{sec:Impact_DL_pilots}
Thus far,  the users have been assumed to be unaware of DL channel estimates, and  the  DL channel statistics  have been used for signal decoding by assuming that  the effective DL channel coefficients can be approximated by their average counterparts by virtue of  channel hardening property of massive MIMO \cite{Marzetta2016_Book}.
For  co-located massive MIMO, channel hardening has been justified  \cite{Ngo2017n}.  Nevertheless, it has been shown in \cite{Ngo2016} that the channel hardening in cell-free massive MIMO occurs only in scenarios where a very large number of APs is deployed in close vicinity. To circumvent this,   in cell-free massive MIMO, the DL channels can be estimated at the users via DL pilots beamformed by APs, and the estimated DL CSI can be used for signal decoding.  Here, we present  a performance analysis of SWIPT  with   DL pilots.

\setcounter{mycounter}{\value{equation}}
\begin{figure*}[!t] 
	\vspace{-7mm}
	\addtocounter{equation}{9}
\begin{eqnarray}\label{eqn:k_WC_Gaussian_Rx_signal_DL}
r_{k,d}   &=&  \underbrace{\sqrt{ \tilde{P_d}}  \E {\left(a_{kk} |\hat{a}_{kk} \right) } q_k }_{\text{desired signal}} + \underbrace{ \sqrt{  \tilde{P_d}} \left(\left(a_{kk} |\hat{a}_{kk} \right) q_k  - \E{\left(a_{kk} |\hat{a}_{kk} \right)  }q_k \right) }_{\text{detection uncertainty}}+ \underbrace{ \sqrt{  \tilde{P_d}} \sum\nolimits_{i \neq k}^{K} \left(a_{ki} |\hat{a}_{kk} \right) q_i }_{\text{Inter-user interference}} + \underbrace{n_{k,d}}_{\text{AWGN}}
\end{eqnarray}
	\vspace{-9mm}
\end{figure*}
\setcounter{equation}{\value{mycounter}}
\setcounter{mycounter}{\value{equation}}
\begin{figure*}[!t] 
	\addtocounter{equation}{15}
	\begin{eqnarray}\label{eqn:rate_DL_up_anlysis}
\E[\hat{a}_{kk}]{\gamma_{k,d}} =   { { \frac{\sum_{m=1}^{M} \sum_{m'=1}^{M} \eta_{mk}^{1/2} \eta_{m'k}^{1/2} \rho_{mk} \rho_{m'k} + \sum_{m=1}^{M} \eta_{mk} \zeta_{mk} \rho_{mk} - \frac{v_{kk}}{\tau_{p,d} P_{p,d} v_{kk} + 1}}{P_d \frac{v_{kk}}{\tau_{p,d} P_{p,d} v_{kk} + 1} +  P_d \sum_{i \neq k}^{K} \sum_{m=1}^{M} \eta_{mi} \zeta_{mk} \rho_{mi} + 1 }}}
\end{eqnarray}
	\vspace{-2mm}
	\hrulefill
\end{figure*}
\setcounter{equation}{\value{mycounter}}

\subsection{Downlink channel estimation at users   }\label{sec:Channel_state_information_acquisition_DL}

Having been locally estimated  UL channels, the APs can now beamform the DL pilots towards the users. This technique guarantees that the DL pilot sequence length depends only on the number of users, and it does not scale with the number of APs \cite{Ngo2016,Ngo2017}.   This technique thus paves a way of efficiently estimating DL channels at the user nodes, while preserving the system  scalability.

Based on the first term of (\ref{eqn:k_th_Us_WC_Gaussian_Rx_signal}), the  effective DL channel coefficient that must be estimated by the  $k$th user  node to decode its desired signal is given by
\begin{eqnarray}\label{eqn:a_ki}
a_{ki} \triangleq \sum_{m=1}^{M} \eta_{mi}^{1/2} h_{mk} \hat{h}_{mi}^*.
\end{eqnarray}
The DL pilot length is $\tau_{p,d}$ in symbol durations (see Fig.\ref{fig:time_fram}b). Thus, an additional $\tau_{p,d}$ duration is needed for DL channels estimation. 

The pilot sequence transmitted by the $m$th AP is denoted by $\boldsymbol{\phi}_{k,d} \in \mathbb C^{ \tau_{p,d} \times 1} $, where $k \in \{1,\cdots,K\}$.
These DL pilots are also  mutually orthogonal,  satisfying $\boldsymbol{\phi}_{k,d}^H \boldsymbol{\phi}_{k',d} = 0 $ for $k \neq k'$ and $\| \boldsymbol{\phi}_{k,d} \|^2 = 1$.
The pilot signal sent by the $m$th AP can be then written as
\begin{eqnarray}\label{eqn:Tx_DL_pilot_vec}
	\mathbf{x}_{pm} = \sqrt{\tau_{p,d} P_{p,d}} \sum_{k=1}^{K} \eta_{mk}^{1/2} \hat{h}_{mk}^* \boldsymbol{\phi}_{k,d} ,
\end{eqnarray}  
where $P_{p,d}$ is the DL pilot transmit power at each AP, and $\hat{h}_{mk}$ is defined in \eqref{eqn:estimate_of_h_mk}. The pilot vector received at the $k$th user can be written as
\begin{eqnarray}\label{eqn:Rx_pilot_vec_K_US}
	{\mathbf{y}'_{k,d}} = \sum_{m=1}^{M} h_{mk} \mathbf{x}_{pm} + {\mathbf{n}'_{pk,d}},
\end{eqnarray} 
where ${\mathbf{n}'_{pk,d}} \sim \mathcal {CN}(\mathbf{0},\mathbf{I})$ is the   $k$th user AWGN vector. The received pilot signal is rewritten by using  (\ref{eqn:Tx_DL_pilot_vec}) and  (\ref{eqn:Rx_pilot_vec_K_US}) as 
\begin{eqnarray}\label{eqn:Rx_pilot_K_US}
	{\mathbf{y}'_{k,d}} = \sqrt{\tau_{p,d} P_{p,d}} \sum_{i=1}^{K} a_{ki} \boldsymbol{\phi}_{i,d}  + {\mathbf{n}'_{pk,d}},
\end{eqnarray}  
where $a_{ki}$ is the effective DL channel and is   defined in (\ref{eqn:a_ki}). 
A sufficient statistic to estimate  this effective DL channel  can be obtained   by projecting \eqref{eqn:Rx_pilot_K_US} into  $\boldsymbol{\phi}_{k,d}$ as 
\begin{eqnarray}\label{eqn:Rx_DL_k}
	y_{k,d} =  \boldsymbol{\phi}_{k,d}^H 	{\mathbf{y}'_{k,d}} = \sqrt{\tau_{p,d} P_{p,d}} a_{kk} + n_{pk,d},
\end{eqnarray}
where $n_{pk,d} = \boldsymbol{\phi}_{k,d}^H {\mathbf{n}'_{pk,d}}  $. By using (\ref{eqn:Rx_DL_k}), the MMSE estimate of the effective channel coefficient (${a}_{kk}$)  is derived  as \cite{Ngo2016}
\begin{eqnarray}\label{eqn:akk_estimate}
	\hat{a}_{kk} &=& \E{a_{kk}} \\
	&&+\frac{\sqrt{\tau_{p,d} P_{p,d}} \Var{a_{kk}} }{\tau_{p,d} P_{p,d} \Var{a_{kk}} + 1} \left(  y_{k,d} - \sqrt{ \tau_{p,d} P_{p,d}} \E{a_{kk}} \right). \nonumber
\end{eqnarray} 
By evaluating $\Var{a_{kk}}$ and $ \E{a_{kk}} $, the MMSE estimate of ${a}_{kk}$   can be derived   as (see Appendix A in \cite{Ngo2016}) 
\begin{eqnarray}\label{eqn:akk_hat}
	\hat{a}_{kk} = \frac{\sqrt{\tau_{p,d} P_{p,d}} v_{kk} y_{k,d} + \sum_{m=1}^{M} \eta_{mk}^{1/2} \rho_{mk}}{\tau_{p,d} P_{p,d} v_{kk} + 1 },
\end{eqnarray}
where $v_{kk}$ is defined as
\begin{eqnarray}\label{eqn:v_kk}
	{v}_{kk} \triangleq \sum_{m=1}^{M} \eta_{mk} \zeta_{mk} \rho_{mk}.
\end{eqnarray}
The actual effective DL channel coefficient  is given by $a_{kk} = \hat{a}_{kk} + \epsilon_{kk}^{a} $, where $\epsilon_{kk}^{a}$ is an  estimation error, which is  independent of $\hat{a}_{kk}$ \cite{Kay1993}.

\subsection{Achievable DL rate with DL pilots/channel estimates}\label{sec:Achievable_DL_rate_DL_pilots}
The received signal at the $k$th user (\ref{eqn:rearranged_kth_Rx_signal}) can  be rewritten by using the effective DL channel coefficient (\ref{eqn:a_ki}) as 
\begin{eqnarray}\label{eqn:Rx_signal_DL}
	r_{k,d} 
	&=& \sqrt{P_d} a_{kk} q_k	+ \sqrt{P_d} \sum_{i \neq k}^{K} a_{ki} q_i + n_{k,d},	
\end{eqnarray}
where $n_{k,d} \sim \mathcal{CN}(0,1)$ is the $k$th user   AWGN.
When  DL channel estimates are available at the $k$th user,  \eqref{eqn:Rx_signal_DL} can be rearranged for signal decoding  as   \eqref{eqn:k_WC_Gaussian_Rx_signal_DL}
\cite{Ngo2016,Medard2000}.
By using \eqref{eqn:k_WC_Gaussian_Rx_signal_DL}, the SINR can be written  as
\addtocounter{equation}{1}
 \begin{eqnarray}\label{eqn:SINR_DL}
 \gamma_{k,d} = \frac{\tilde{P_d}|\E{a_{kk} | \hat{a}_{kk}} |^2 }{\tilde{P_d}\displaystyle \sum_{i=1}^{K} \E{|a_{ki}|^2 | \hat{a}_{kk}}  - \tilde{P_d} |\E{a_{kk}| \hat{a}_{kk}} |^2 + 1 }.
 \end{eqnarray}	 
By noticing that   $a_{kk}$ is Gaussian distributed,  and the pairs $(\hat{a}_{kk},\epsilon_{kk}^{a})$  and $(a_{kk},a_{ki})$ are independent for $k \neq i$, the SINR in \eqref{eqn:SINR_DL} can be rewritten as follows:
	 \begin{eqnarray}\label{eqn:SINR_DL_reduced}
	 \gamma_{k,d} = \frac{\tilde{P_d} | \hat{a}_{kk} |^2 }{\tilde{P_d} \sum_{i \neq k}^{K} \E{|a_{ki}|^2}+ \tilde{P_d} \E{|\epsilon_{kk}^{a} |^2 }    + 1 }.
	 \end{eqnarray}
Then, the achievable DL rate can be defined as
\begin{eqnarray}\label{eqn:rate_DL}
	R_{k,d} =  \frac{\left(1-\alpha\right) \tau_{d,d}}{\tau_{c}}\E[\hat{a}_{kk}]{\log[2] {1+ \gamma_{k,d}} },
\end{eqnarray}
where $\tau_{d,d} = \tau_{c}-(\tau_{p} + \tau_{p,d} + \tau_{u})$.  By invoking Jensen's inequality, an upper bound of the DL rate at the $k$th user node can be derived as
\begin{eqnarray}\label{eqn:rate_DL_up}
	R_{k,d}^{ub} =  \frac{ \left(1-\alpha\right)\tau_{d,d}}{\tau_{c}} {\log[2] {1+ \E[\hat{a}_{kk}]{\gamma_{k,d}}} },
\end{eqnarray}
where $\E[\hat{a}_{kk}]{\gamma_{k,d}}$ is given by 
\begin{eqnarray}\label{eqn:rate_DL_up_full}
	 \E[\hat{a}_{kk}]{\gamma_{k,d}} =    { { \frac{\tilde{P_d} \E{| \hat{a}_{kk} |^2} }{\tilde{P_d}  \E{|\epsilon_{kk}^{a} |^2 }  + \tilde{P_d} \sum_{i \neq k}^{K}  \E{|a_{ki}|^2} +1 } } }.
\end{eqnarray}
 By evaluating expectation terms in \eqref{eqn:rate_DL_up_full}, 
 the effective SINR can be written as in \eqref{eqn:rate_DL_up_anlysis} (see Appendix \ref{app:Appendix4} for derivation).

\setcounter{mycounter}{\value{equation}}
\begin{figure*}[!t] 
	\vspace{-7mm}
	\addtocounter{equation}{1}
	\begin{eqnarray}
	F_{\mathcal R_{\min}}(\mathcal R) &=&\mathrm{P_r}\left(\mathcal R_{\min} =  \min_{k} \left(\frac{\left(1-\alpha\right) \tau_d}{\tau_c} \log[2] {1+ \gamma_k}\right) \leq \mathcal{R} \right)  \label{eqn:DCF_rate_HE1} \\
	F_{\mathcal E_{\min}}(\mathcal E)&=& \mathrm{P_r}\left(\mathcal{E}_{\min} = \min_{k}\left(\alpha \tau_d \Psi_{EH} \left( {P_k} \right) +  (1-\alpha) \tau_d \Psi_{EH} \left(\theta { P_k}\right)\right) \leq \mathcal{E} \right) \label{eqn:DCF_rate_HE2}
	\end{eqnarray}
	\vspace{-2mm}
	\hrulefill
\end{figure*}
\setcounter{equation}{\value{mycounter}}
\section{Numerical Results}\label{sec:Numerical}                                       
  The path-loss between the $m$th AP and the $k$th user is modeled as  $\zeta_{mk} = (d_0/d_{mk})^{\nu} \times 10^{\varphi_{mk}/10}$,
where $d_{mk}$ is the   distance between the $m$th AP and the $k$th user, $d_0$ is the reference distance,  $\nu$ is the path-loss exponent, and $10^{\varphi_{mk}/10}$ captures log-normal shadow fading with $\varphi_{mk} \sim \mathcal{N}(0,\sigma_{sh}^2)$ \cite{Marzetta2016_Book}. The APs are uniformly distributed, while the user nodes are randomly distributed over an area of $D \times D \,\text{m}^{2}$. Other simulation parameters are tabulated in Table \ref{tab:sim_para}.

\begin{table} [!h]
	\caption{System parameters for the simulations}
	\vspace{-5mm}
	\begin{center}\label{tab:sim_para}
		\begin{tabular}{|c|c|}
			\hline
			Parameter & value  \\
			\hline
			 Carrier frequency & 2\,GHz  \\
			\hline
			 Bandwidth &  20\,MHz \\
			\hline
			 Noise figure & 10\,dB  \\
			\hline
		 	Noise variance & -91\,dBm  \\
			\hline
			$D$,\,$d_0$ & 300\,m, 1\,m  \\
			\hline
		 	$\sigma_{sh}^2$ & 8\,dB  \\
			\hline
		 	$\tau_c$, $\tau_p = \tau_{p,d}$ & 196, $K$  \\
			\hline
			$\nu$  & 3.4 \\
			\hline
		\end{tabular}
	\end{center} 
\vspace{-5mm}
\end{table}

\begin{figure}[!t]\centering\vspace{-1mm}
	\includegraphics[width=0.4\textwidth]{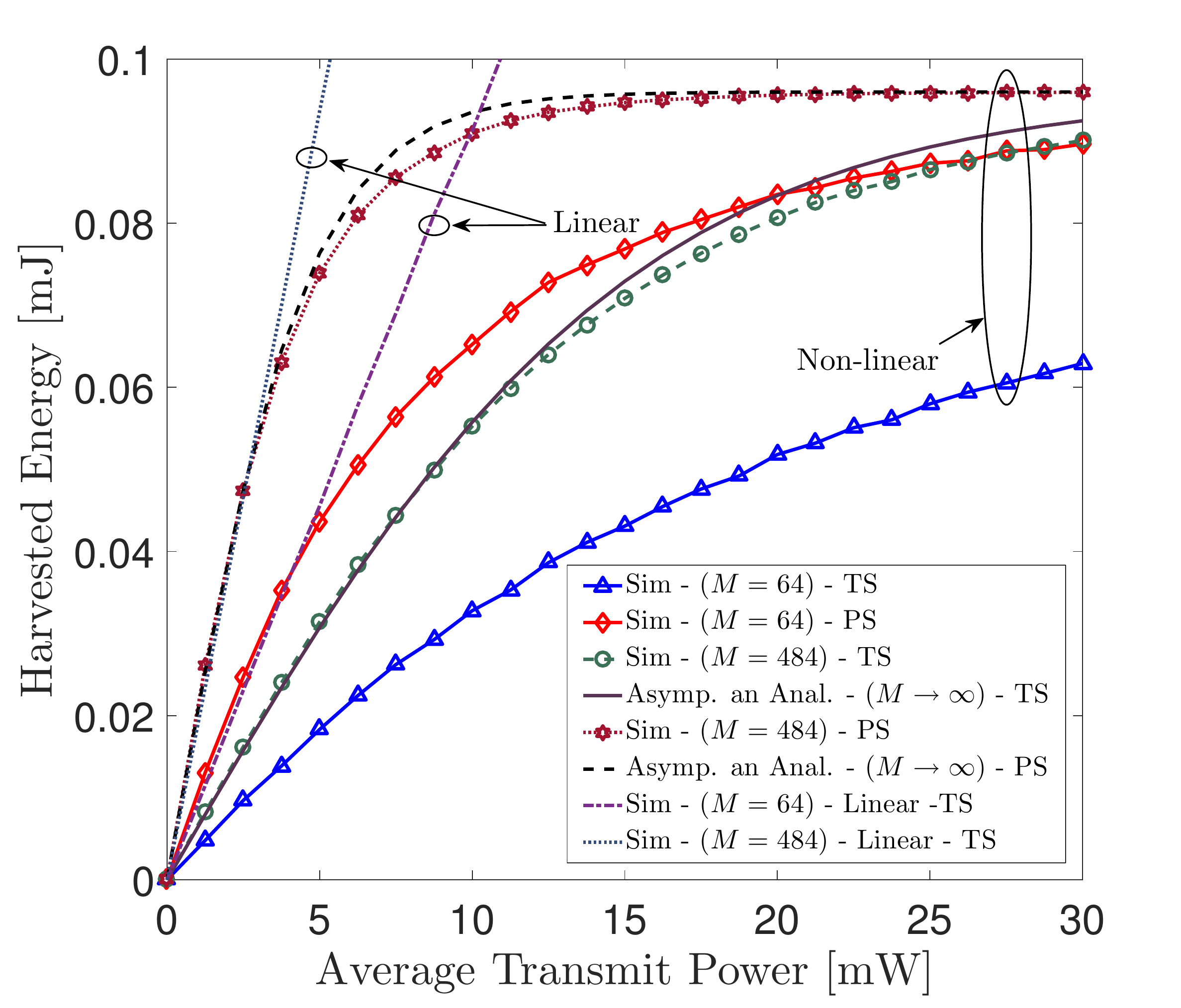}\vspace{-5mm}
	\caption{ The harvested energy versus the average transmit power for TS and PS protocols for $\alpha = 0.5$, $\theta = 0.5$. }
	\label{fig:HE_VS_TxPower_1_US} \vspace{-5mm}
\end{figure}

In Fig. \ref{fig:HE_VS_TxPower_1_US}, the harvested energy curves for    TS and PS protocols are plotted against the average transmit power ($\bar{\gamma}$) by using linear and non-linear SWIPT models. 
The curves for the finite and infinite AP regimes are plotted via Monte-Carlo simulations and our asymptotic analysis \eqref{eqn:asymptotic_Rx_power}, respectively. 
The harvested energy of non-linear SWIPT model (\ref{eqn:EH_model}) is compared to that of  the   linear model with a fixed RF-to-DC conversion efficiency of 0.9. Fig. \ref{fig:HE_VS_TxPower_1_US} reveals that the linear SWIPT model overwhelmingly overestimates the harvested energy with respect to the non-linear counterpart in medium-to-high average transmit power regime. 
Fig. \ref{fig:HE_VS_TxPower_1_US} also depicts that the harvested energy can be boosted  by increasing the  density of   distributed APs in a given geographical area. For instance with PS protocol, at an average transmit power of 5\,mW, the harvested energy can be improved by $69.4$\% when the number of APs is increased from 64 to 484.
The PS  protocol outperforms TS protocol because the former is able to perform energy harvesting during the entire DL time-slot via power-splitters, while the latter employs time-division technique, which is typically sub-optimal.

\begin{figure}[!t]\centering\vspace{-0mm}
	\includegraphics[width=0.4\textwidth]{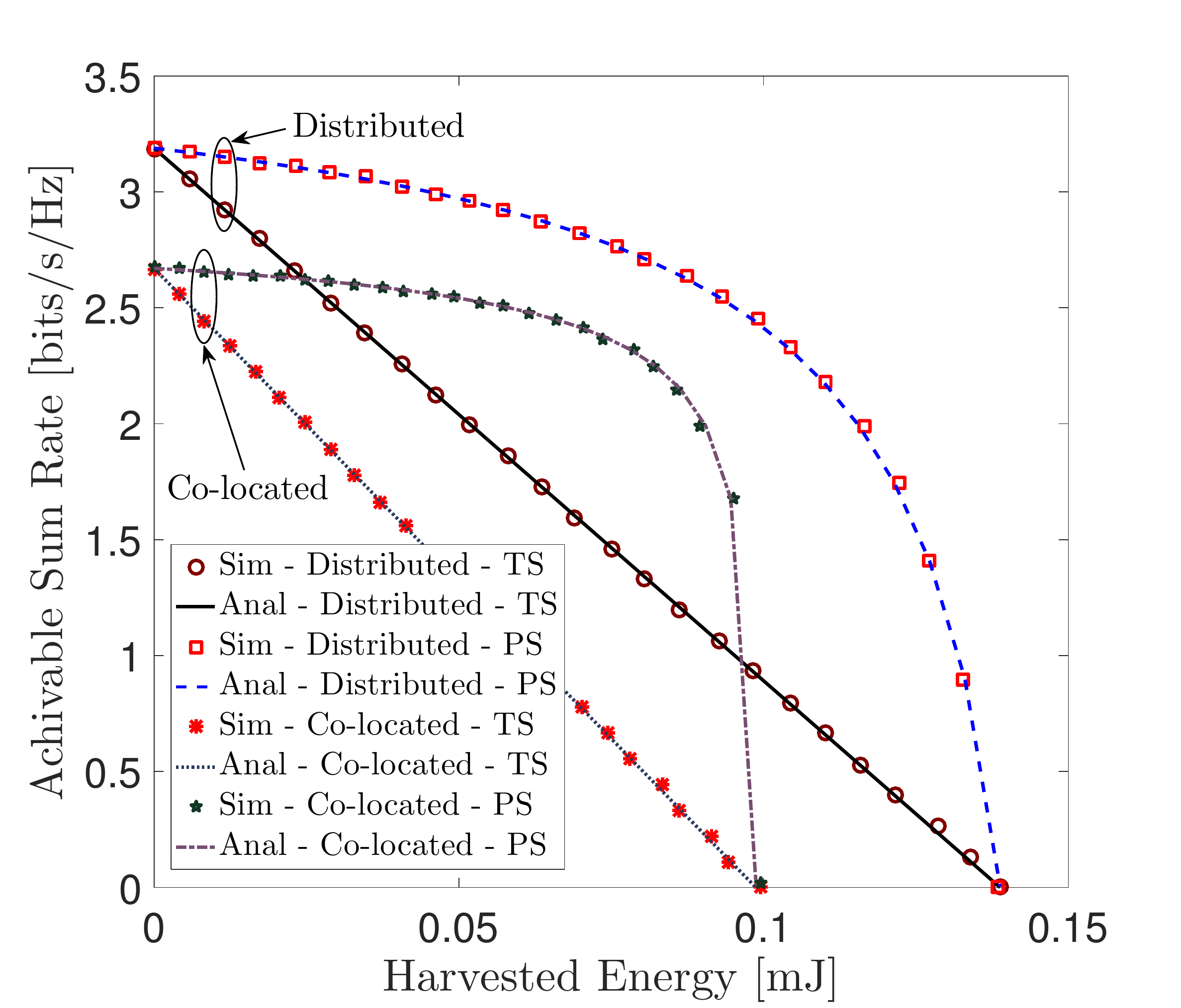}\vspace{-5mm}
	\caption{ A comparison of the energy-rate trade-off between the distributed and co-located massive MIMO architectures of TS and PS protocols for $M = 64$, $K = 2$ and $\bar{\gamma} = 10$\,dBm. }
	\label{fig:equal_power_comparison} \vspace{-5mm}
\end{figure} 

In Fig. \ref{fig:equal_power_comparison}, the energy-rate trade-offs of co-located and cell-free massive MIMO are compared.  
In the former case, the BS is equipped with an $M=64$ antenna array, while the same number of APs is uniformly distributed in the latter case. User locations are  fixed for both systems.
These trade-off curves are plotted by using \eqref{eqn:average_HE} and \eqref{eqn:kth_US_rate} with $\alpha = 0$ and $\theta=0$ for PS and TS protocols, respectively. Fig. \ref{fig:equal_power_comparison} confirms that the distributed 
APs outperform the co-located counterpart. For example, at a sum rate of $1.5$\,bits/s/Hz, the cell-free massive MIMO provides   harvested energy gains of $71.4$\% and  $33.7$\%, respectively, for TS and PS protocols,  over the co-located massive MIMO. 
However, cell-free set-up achieves this performance gain at the expense of an additional backhaul/fronthaul requirement. 
Moreover, Fig. \ref{fig:equal_power_comparison} depicts the fundamental trade-off between the harvested energy and the  sum rate. On one hand, 
the sum rate is maximized when the both TS and PS factors approach zero, and at this operating point, the entire time-slot allocated for the DL is  used only for data transfer. Simultaneously,   the harvested energies at the users become infinitesimal.  On the other hand, the harvested energy becomes a maximum when the     TS and PS factors approach unity, and consequently at this operating point, the  sum rate vanishes. Thus, by traversing through the energy-rate trade-off curves in Fig. \ref{fig:equal_power_comparison}, the harvested energies and  sum rates can be obtained for any given    TS/PS factor.

\begin{figure}[!t]\centering\vspace{-5mm}
	\includegraphics[width=0.4\textwidth]{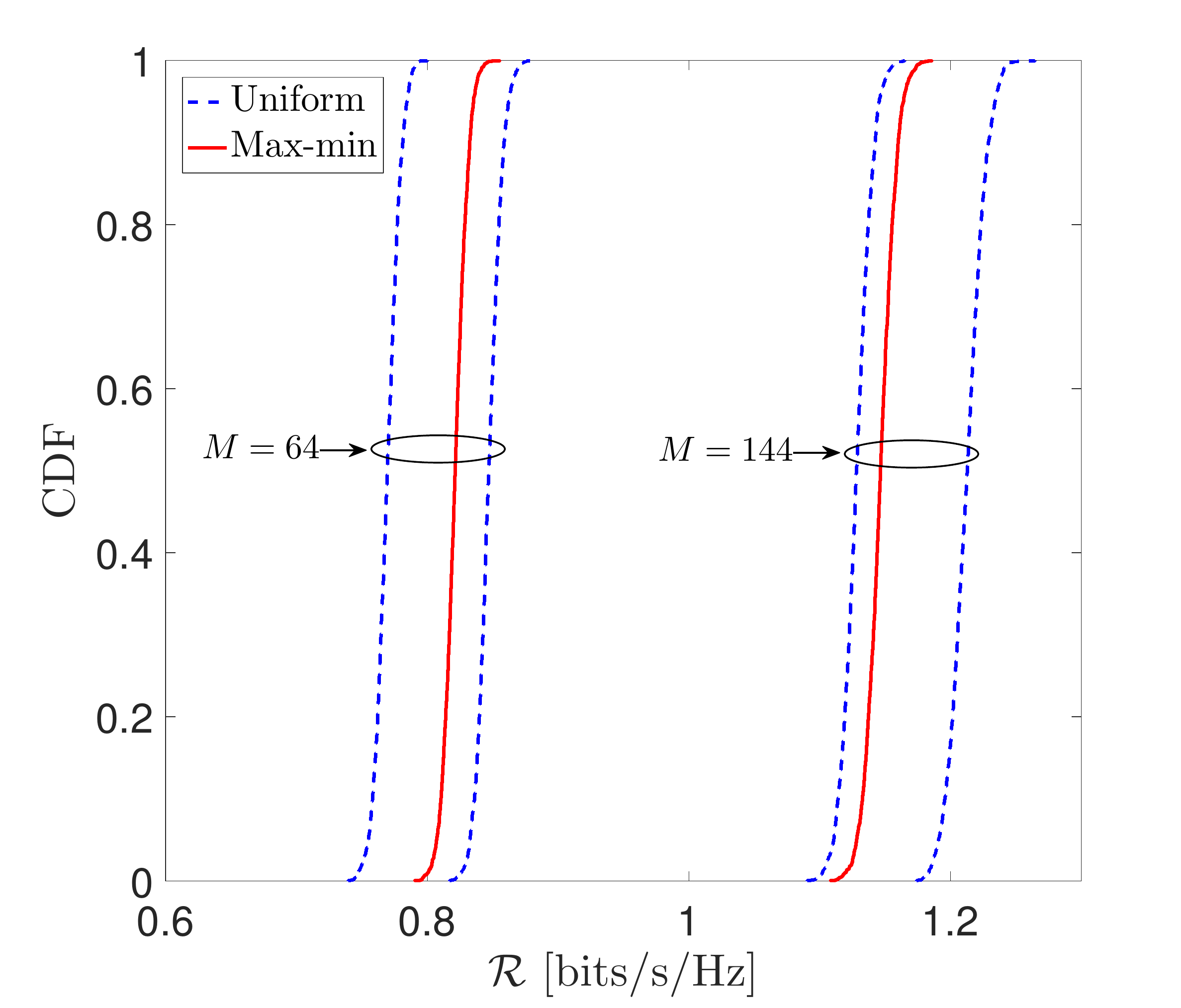}\vspace{-5mm}
	\caption{The CDF of the minimum achievable user rate of PS protocol for $K=2$, $\bar{\gamma} = 10$\,dBm and $\theta = 0.5$. }
	\label{fig:energy_rate_trade_off_ps_rate_2US} \vspace{-4mm}
\end{figure}

\begin{figure}[!t]\centering\vspace{-0mm}
	\includegraphics[width=0.4\textwidth]{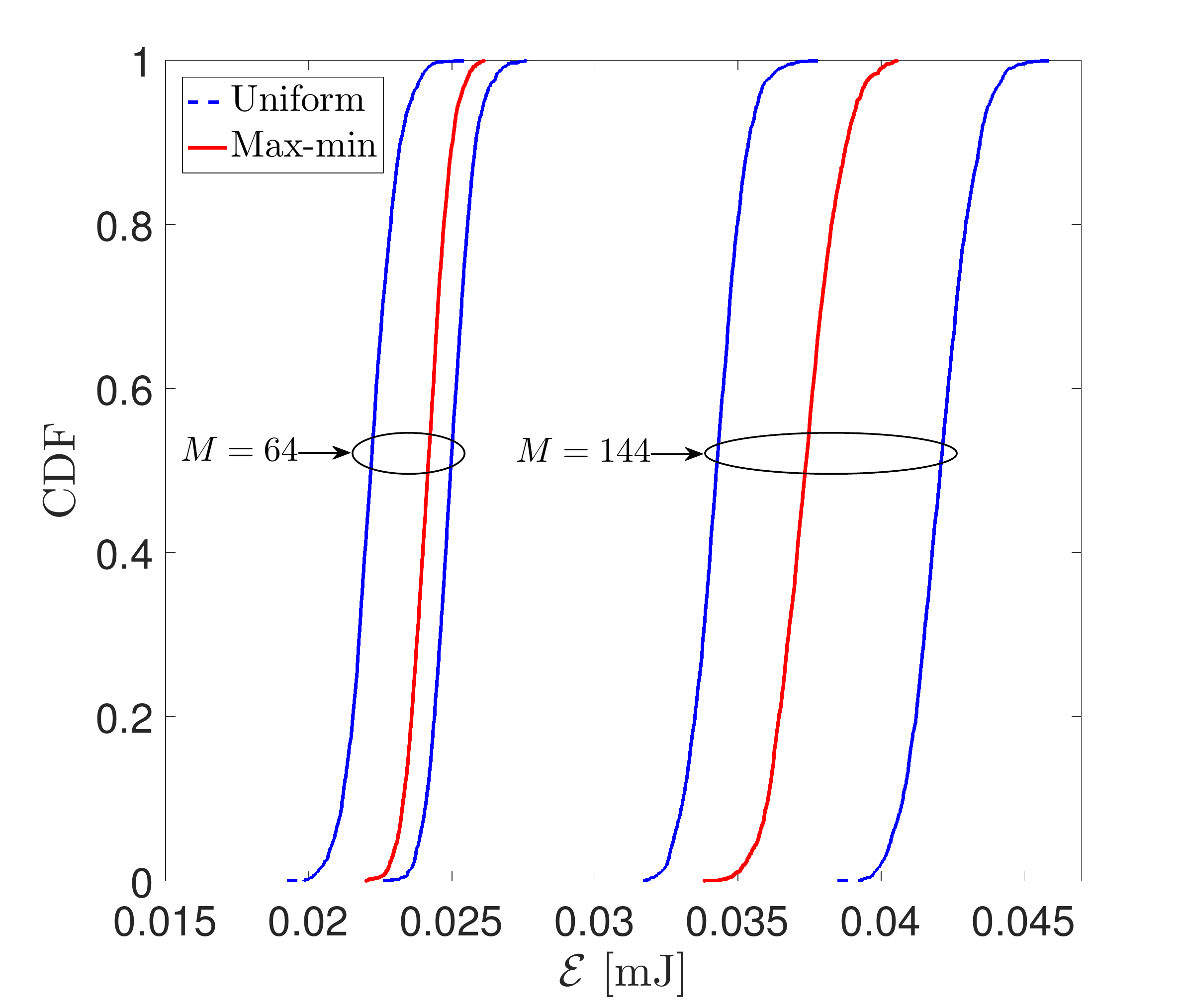}\vspace{-5mm}
	\caption{The CDF of the minimum user harvested energy of PS protocol for $K=2$, $\bar{\gamma} = 10$\,dBm and $\theta = 0.5$. }
	\label{fig:energy_rate_trade_off_ps_energy_2US}   \vspace{-5mm}
\end{figure}

In    Fig. \ref{fig:energy_rate_trade_off_ps_rate_2US} and Fig. \ref{fig:energy_rate_trade_off_ps_energy_2US},  the performance of 
the proposed max-min fairness-based transmit power control is compared against the uniform power control. To this end, the cumulative distribution functions (CDFs) of  the minimum  user rate and minimum user harvested energy    pertaining to PS protocol are plotted in Fig. \ref{fig:energy_rate_trade_off_ps_rate_2US} and Fig. \ref{fig:energy_rate_trade_off_ps_energy_2US}, respectively, by varying the number of APs as  $M=64$ and $M=144$.  
	These CDFs are evaluated as given in \eqref{eqn:DCF_rate_HE1} and \eqref{eqn:DCF_rate_HE2},
 where $\mathrm{P_r}(\cdot)$ denotes the probability operator.
This comparison   reveals the performance gains that can be acquired by adopting  the proposed max-min power control over the uniform counterpart. For instance as per Fig. \ref{fig:energy_rate_trade_off_ps_rate_2US}, at $M=64$, the 90\%-likely  user rate of the max-min  power control is about $0.807$\,bits/s/Hz, which is in turn $1.08$ times higher than  that of the uniform power control, which provides a minimum user rate about  $0.757$\,bits/s/Hz at the same operating point. 
Similarly, Fig. \ref{fig:energy_rate_trade_off_ps_energy_2US} reveals that the max-min power control renders a  $1.09$ times gain in terms of the 90\%-likely harvested energy over the uniform counterpart at $M=64$. Fig. \ref{fig:energy_rate_trade_off_ps_rate_2US} and Fig. \ref{fig:energy_rate_trade_off_ps_energy_2US} also reveal that  the 90\%-likely  user rate and harvested energy can be boosted by $1.41$ and $1.56$ times, respectively, when the number of APs is increased from $M=64$ to $M=144$. Thus, the proposed max-min power control  guarantees to enhance the user rate and harvested energy   with weak channel conditions. 

\begin{figure}[!t]\centering\vspace{-5mm}
	\includegraphics[width=0.4\textwidth]{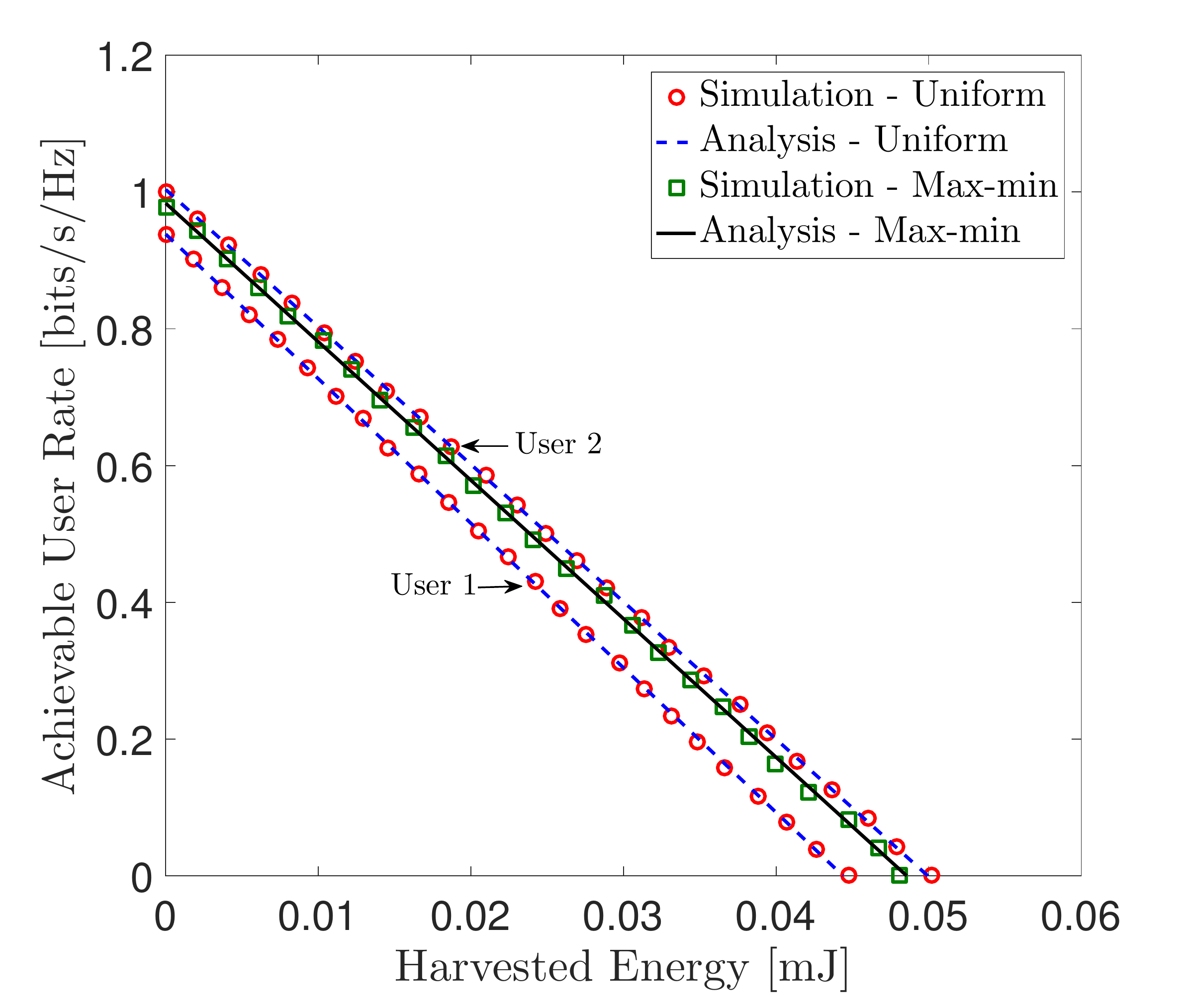}\vspace{-5mm}
	\caption{ The energy-rate trade-off of TS protocol  for  $K = 2$ and $\bar{\gamma} = 10$\,dBm, and  $M = 64$. }
	\label{fig:trade_off_power_allocation_comparison_final}   \vspace{-5mm}
\end{figure}

In Fig. \ref{fig:trade_off_power_allocation_comparison_final}, the achievable energy-rate trade-off of the TS protocol is investigated. 
By using the optimal solutions of the  proposed max-min transmit power control policies in  section \ref{sec:Transmit_power_control_of_rate_for_TS_protocol}, the max-min fairness optimal/common energy-rate trade-off (\ref{eqn:TS_trade_off}) is plotted. 
The  noise at the $k$th user's receiver is modeled as AWGN\footnote{The noise variance in dBm, $\sigma_{n}^2   =  -174 + 10 \log[2]{B} + N_f$, where  $B = 20$\,MHz is the channel bandwidth and $N_f = 10$\,dB is the  noise figure of the receiver.} with variance, $\sigma_n^2 = -91$\,dBm.
The energy-rate trade-offs of individual users with a uniform power control policy  are also plotted for comparison purposes.  Fig. \ref{fig:trade_off_power_allocation_comparison_final} reveals that users experience distinct energy-rate trade-offs when the uniform power control is adopted, and thus, the underlying performance metrics are dependent on the  detrimental near-far effects.  However, once the proposed max-min power control is employed, all users achieve a common energy-rate trade-off regardless of   near-far effects. 

\begin{figure}[!t]\centering\vspace{-0mm}
	\includegraphics[width=0.4\textwidth]{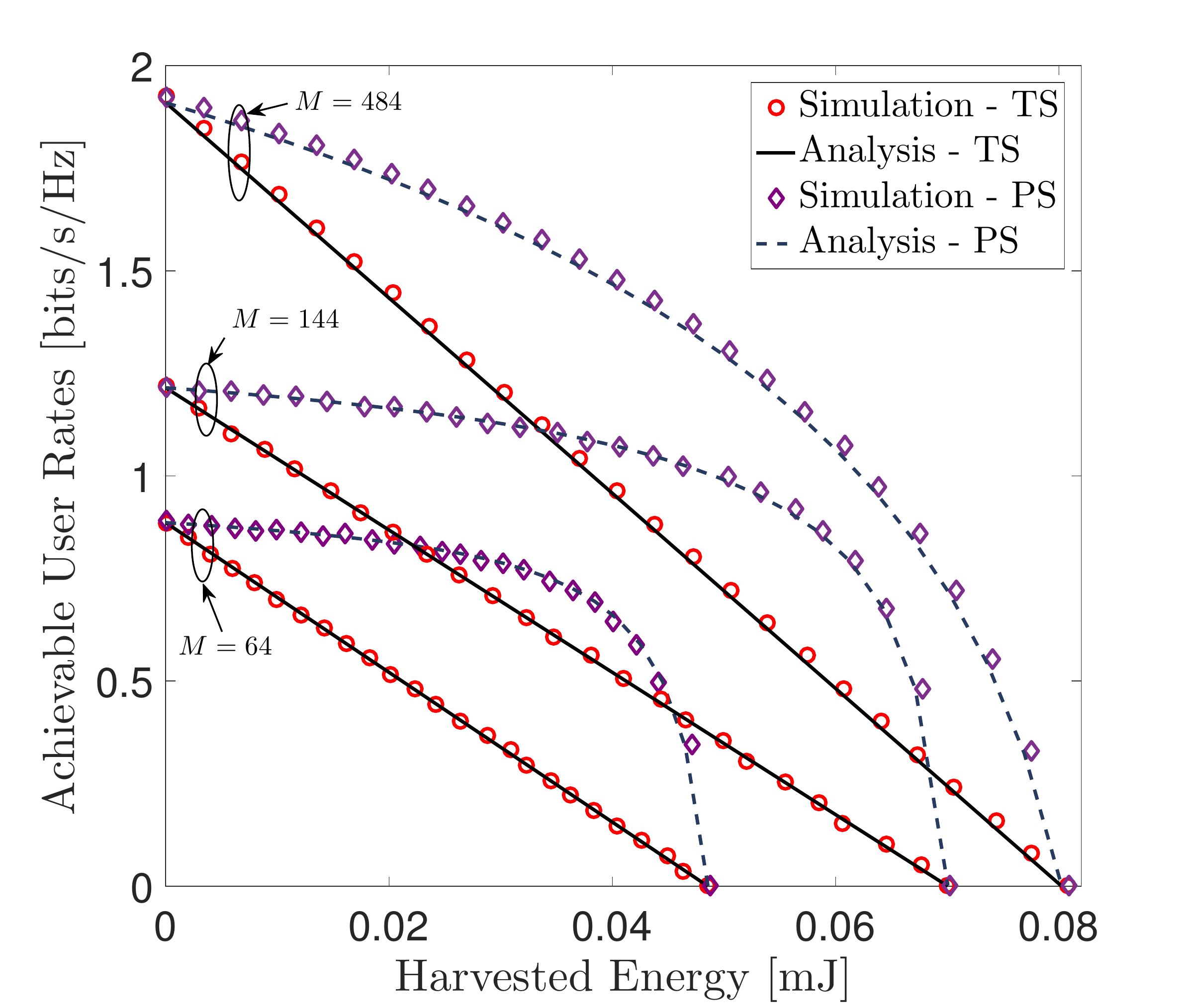}\vspace{-5mm}
	\caption{ A comparison of the energy-rate trade-off with different number of APs for  $K = 2$ and $\bar{\gamma} = 10$\,dBm.. }
	\label{fig:trade_off_finite_asym_64_144_484_final}   \vspace{-5mm}
\end{figure} 

In Fig. \ref{fig:trade_off_finite_asym_64_144_484_final},
the impacts of number of APs on the   energy-rate trade-offs of TS and PS protocols are studied. Three sets of trade-off curves are plotted by changing the number of APs as $M\in\{64, 144, 484\}$. We observe from Fig. \ref{fig:trade_off_finite_asym_64_144_484_final} that the energy-rate trade-off can be boosted significantly by increasing the AP density in a given geographical area. For instance, at a harvested energy of $0.02$\,mJ, the system with $M=144$ APs achieves a sum rate gain of $64$\% over the system with $M=64$ for TS protocol.
The asymptotic ($M\rightarrow \infty$) energy-rate trade-off curves are plotted by using \eqref{eqn:asymptotic_HE} and \eqref{eqn:asymptotic_rate}. The Monte-Carlo simulation with $M=484$ matches well with the asymptotic analysis in \eqref{eqn:asymptotic_HE} and \eqref{eqn:asymptotic_rate}. Thus, the asymptotic performance limits can be obtained in practice with large but finite AP regime. 

\begin{figure}[!t]\centering\vspace{-5mm}
	\includegraphics[width=0.4\textwidth]{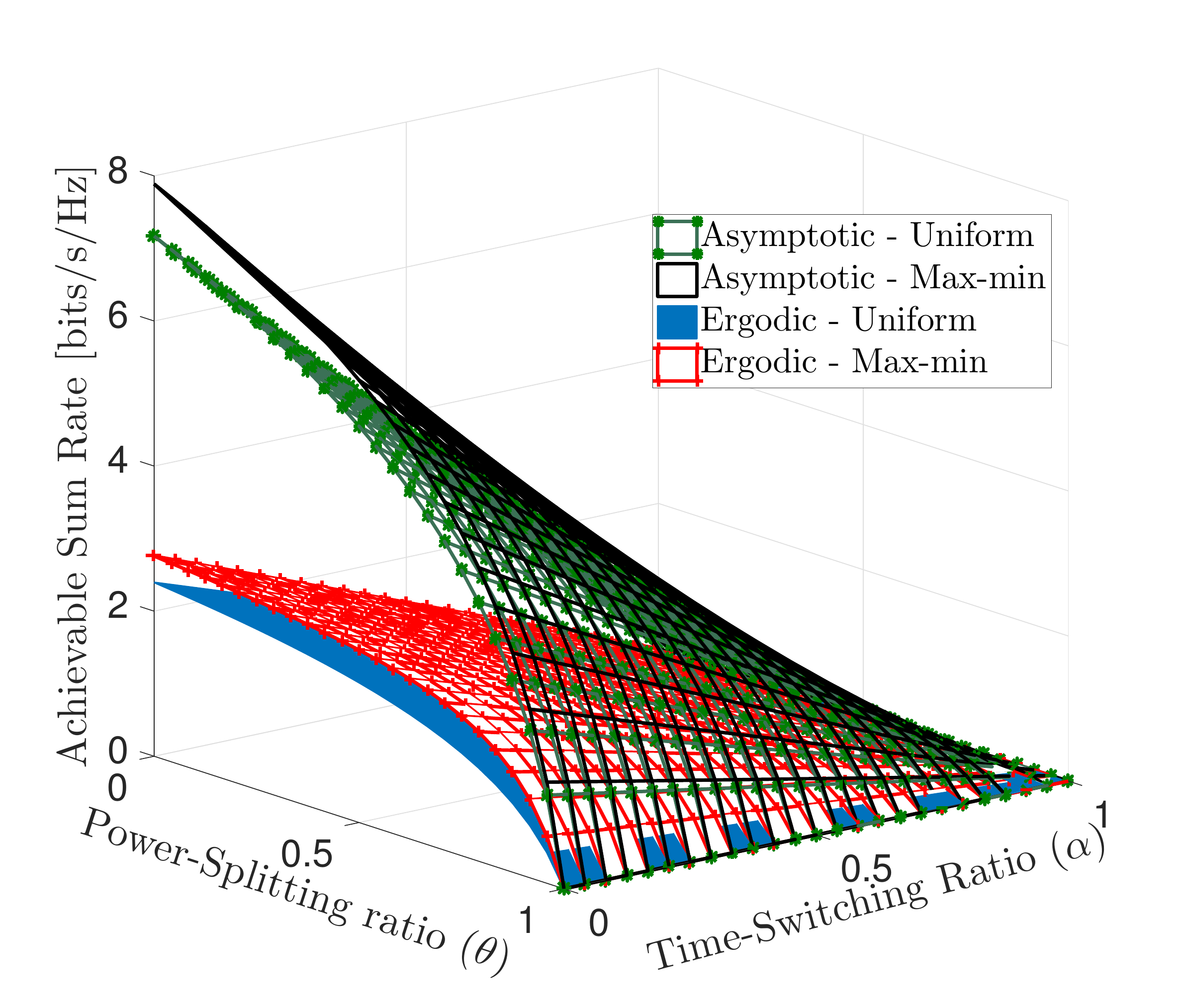}\vspace{-5mm}
	\caption{ The achievable sum rate as a function of the TS and PS factors for for $K = 2$ and $\bar{\gamma} = 10$\,dBm. }
	\label{fig:rate_alpha_theta_final}   \vspace{-5mm}
\end{figure}

\begin{figure}[!t]\centering\vspace{-0mm}
	\includegraphics[width=0.4\textwidth]{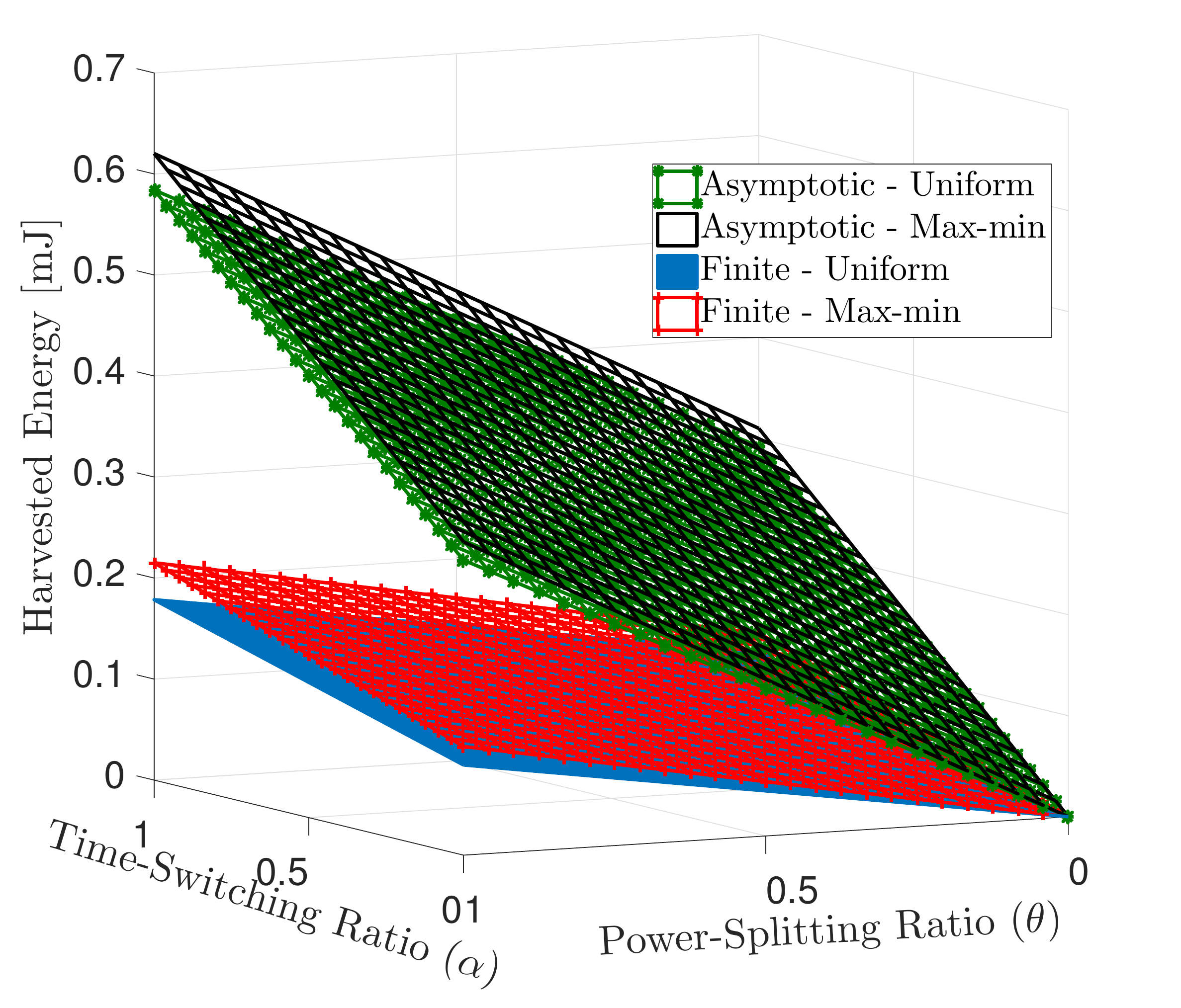}\vspace{-5mm}
	\caption{ The total harvested energy as a function of the TS and PS factors for for $K = 2$ and $\bar{\gamma} = 10$\,dBm.}
	\label{fig:energy_alpha_theta_final}   \vspace{-5mm}
\end{figure}

In Fig. \ref{fig:rate_alpha_theta_final} and Fig. \ref{fig:energy_alpha_theta_final}, the joint effects of the hybrid TS and PS protocol  are investigated by plotting the sum rate \eqref{eqn:kth_US_rate} and harvested energy \eqref{eqn:average_HE}, respectively, as a function of  TS  ($\alpha$) and PS  ($\theta$) factors. The max-min fairness optimal  rate for PS protocol is plotted by adopting the optimal solution of \eqref{eqn:PS_power_control_SNR_1}. Moreover, for PS protocol,  the max-min optimization for the harvested energy  \eqref{eqn:PS_power_control_HE} is adopted. When both $\alpha$ and $\theta$ approach unity, the  sum rate becomes infinitesimal (see Fig. \ref{fig:rate_alpha_theta_final}), whereas the harvested energy becomes a maximum (see Fig. \ref{fig:energy_alpha_theta_final}). Moreover, Fig. \ref{fig:rate_alpha_theta_final} and Fig. \ref{fig:energy_alpha_theta_final} reveal   impact of  the number of APs on the corresponding performance metrics. 

\begin{figure}[!t]\centering\vspace{-5mm}
	\includegraphics[width=0.4\textwidth]{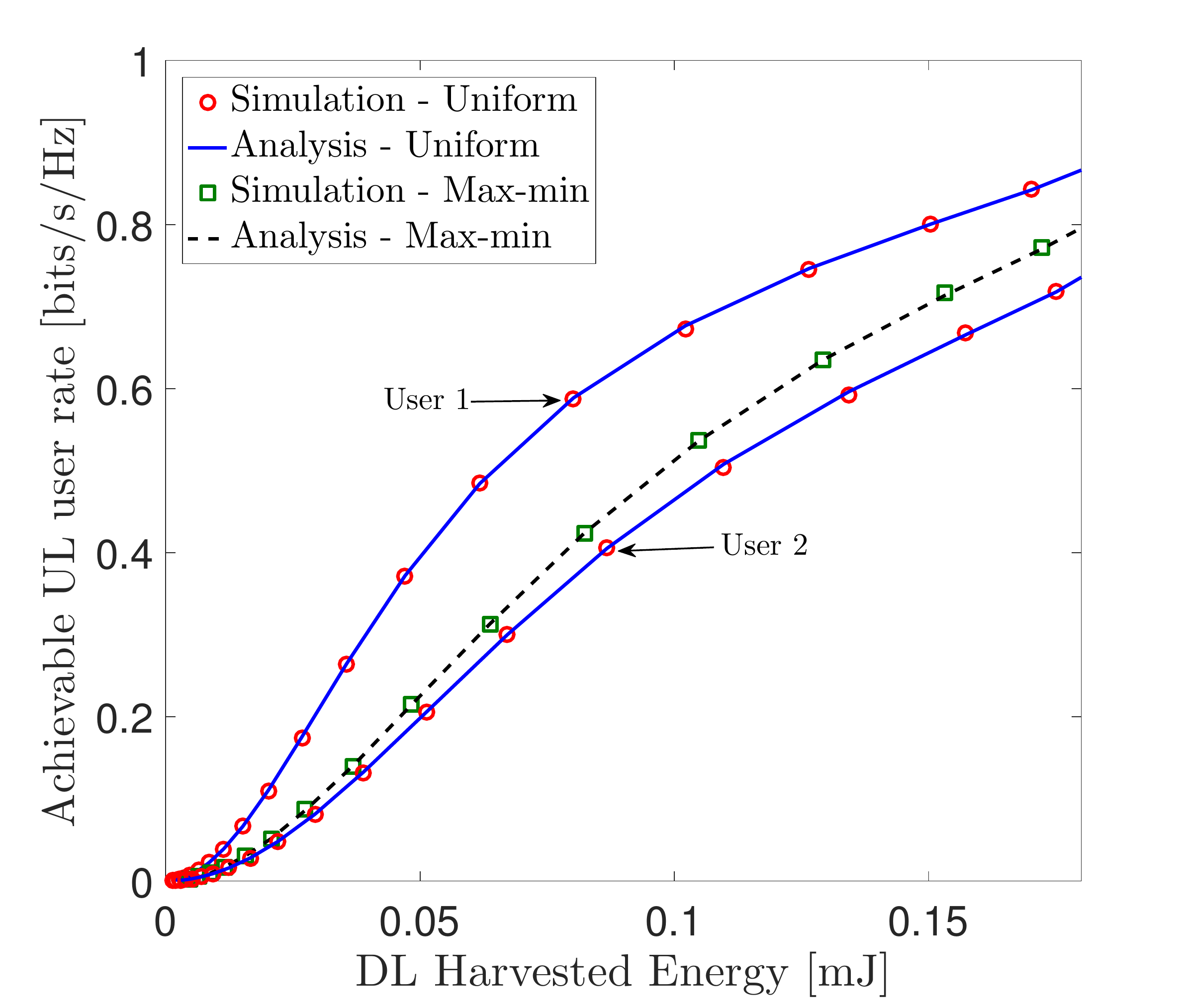}\vspace{-5mm}
	\caption{ Achievable UL user rate versus DL harvested energy for $M = 64$, $K = 2$,  $\alpha = 0.5$ and $\theta = 0.5$. }
	\label{fig:UL_2US_rates}   \vspace{-4mm}
\end{figure}

In Fig. \ref{fig:UL_2US_rates}, the performance of the proposed joint DL/UL transmit power control is investigated by 		plotting 
the   UL user rate  as a function of the DL harvested energy by varying the average DL transmit power. The harvested energy in the DL (i.e., \eqref{eqn:p_uk} with $\kappa = 0.85$) is used for UL transmission. Specifically, 
the DL harvested energy versus UL  rate trade-off of the proposed max-min fairness based joint transmit power control is compared with that of the uniform counterpart.  
The max-min optimal curve is plotted by solving the optimization problem in \eqref{eqn:power_control_UL_SNR}. Fig. \ref{fig:UL_2US_rates} shows that
the joint max-min power control yields a common trade-off between the DL energy and UL rate regardless of the distributed nature of APs/users and thus, preserving the user-fairness in terms of the DL energy harvesting and   UL user rates. However, the   same trade-off is adversely effected when the uniform power control is adopted owing to the inherent near-far effects in a cell-free massive MIMO set-up.    

\begin{figure}[!t]\centering\vspace{-0mm}
	\includegraphics[width=0.4\textwidth]{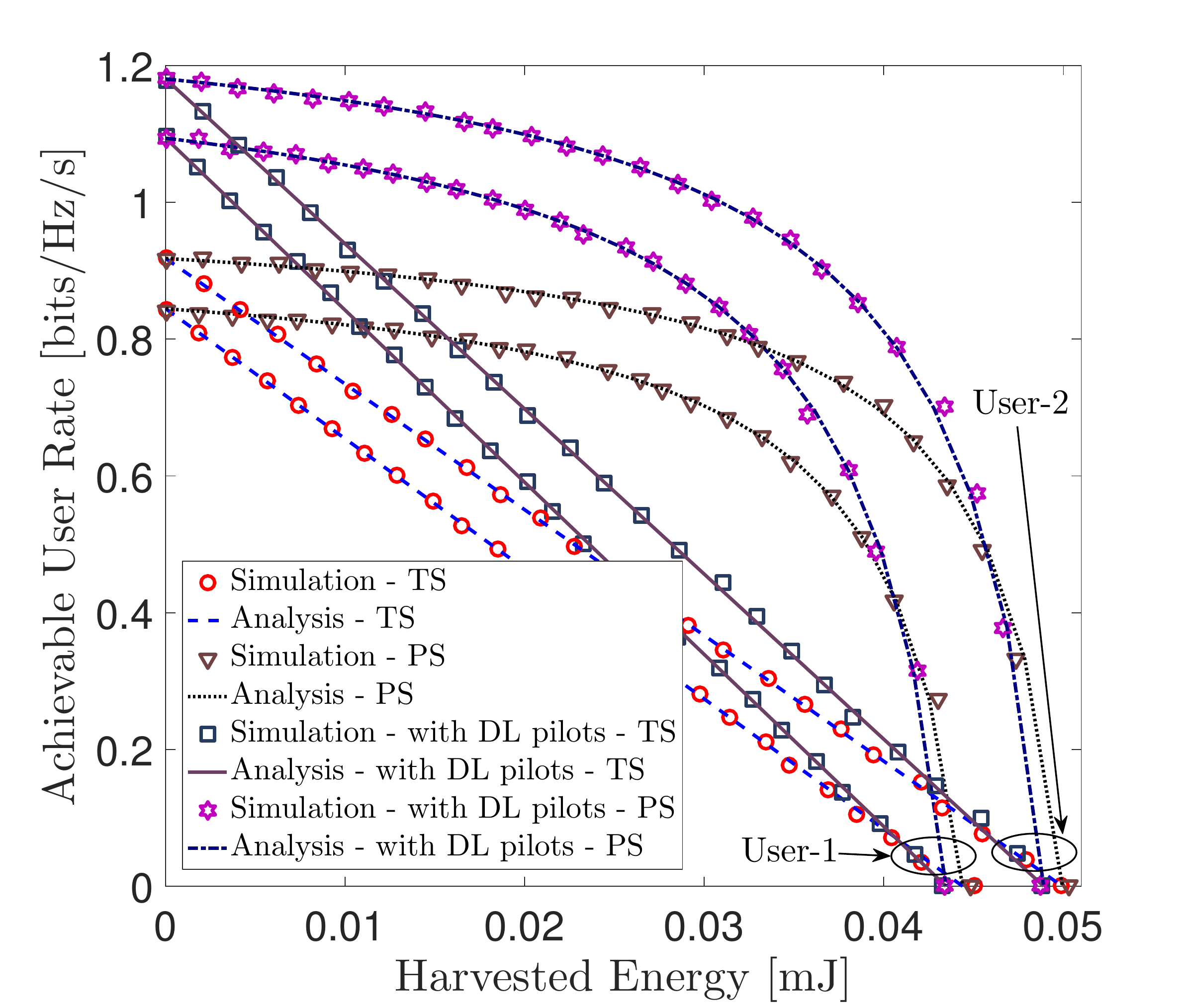}\vspace{-5mm}
	\caption{A comparison of the energy-rate trade-off with/without DL pilots for $M = 64$, $K = 2$, and $\bar{\gamma} = 10$\,dBm.}
	\label{fig:rate_energy_trade_off_comparison_DL_pilots_2_US}   \vspace{-5mm}
\end{figure}

In Fig. \ref{fig:rate_energy_trade_off_comparison_DL_pilots_2_US}, 
the impact of DL pilots  on the   energy-rate trade-off is investigated for both TS and PS protocols by using Monte-Carlo simulations and our analysis in \eqref{eqn:rate_DL_up_full} and \eqref{eqn:rate_DL_up_anlysis}.   Fig. \ref{fig:rate_energy_trade_off_comparison_DL_pilots_2_US} clearly revels that when the estimated CSI, which is facilitated by beamforming DL pilots, is utilized at the users for signal decoding instead of relying on statistical CSI, the achievable rate can be boosted substantially. 
Since the task of beamforming  DL pilots requires an additional time-slot of length $\tau_{p,d}$, which is otherwise be used for energy harvesting purposes, the   harvested energy   
is  affected detrimentally. Thus,  Fig. \ref{fig:rate_energy_trade_off_comparison_DL_pilots_2_US}  shows that   the harvested energy of  the case with DL pilots is slightly less than that of the case without DL pilots.
Nevertheless, as per Fig. \ref{fig:rate_energy_trade_off_comparison_DL_pilots_2_US},  the user rate gain acquired through enabling estimated DL CSI at users via beamforming DL pilots is more prominent than the slight degradation of the harvested energy.  Thus, the use of DL pilots is useful in enhancing the overall energy-rate trade-off of SWIPT operated in a cell-free massive MIMO set-up. 

\section{Conclusion}\label{sec:conclusion}

The benefits of leveraging cell-free massive MIMO in practically realizing   SWIPT have been investigated.  
The feasibility of utilizing a portion of DL harvested energy in UL user transmissions has been  explored  by deriving the  UL user rates. 
By adopting a non-linear energy harvesting model, locally estimated UL CSI at the APs and statistical CSI at the users, tight approximations to the  harvested energies, achievable user rates, and energy-rate trade-offs have been derived. The impact of   DL channel estimation at users via DL pilot beamfoming to boost the   energy-rate trade-off has been studied.
We reveal that user fairness in terms of both harvested energy and   rates can be guaranteed by adopting  max-min based transmit power control policies. By deriving the max-min   optimal energy-rate trade-offs, we reveal that a common harvested energy and user rate at any given TS/PS factor can be achieved for all users regardless of   near-far effects. By deriving max-min optimal power control coefficients for the UL,  we show that the detrimental effects of near-far effect on the UL   user rates can also be mitigated.   We reveal that the   linear energy harvesting model   overestimates the SWIPT performance   over those of the non-linear counterpart. 
Through a max-min   energy-rate trade-off  comparison, we conclude that a cell-free massive MIMO set-up can potentially boost the SWIPT performance over a co-located   counterpart at the expense of additional backhaul/fronthaul requirements. 

\appendices

\section{Derivation of $\E{P_k}$ in \eqref{eqn:avg_power_kth_Us}} \label{app:Appendix2}

By substituting \eqref{eqn:estimate_of_h_mk} and \eqref{eqn:instant_Rx_power_k_user} into \eqref{eqn:avg_power_kth_Us}, the average received power can be derived as 
\addtocounter{equation}{3}
\begin{eqnarray}\label{eqn:identity_1_2}
 	\E{P_k} &=& P_d \E{\left|\sum_{m=1}^{M} \eta_{mk}^{1/2} h_{mk} \hat{h}_{mk}^* \right|^2 }   \\
 	&&\!\!\!\!\! \!\!\!\!\! \!\!\!\!\! +P_d \E{\left|\sum_{m=1}^{M} \sum_{i \neq k}^{K} \eta_{mi}^{1/2} h_{mk} \hat{h}_{mi}^* \right|^2 } \nonumber \\
 	&&\!\!\!\!\!\! \!\!\!\!\! \!\!\!\!\! = P_d \sum_{m=1}^{M} \eta_{mk} c_{mk}^2 \zeta_{mk} \left(2 \tau_p P_p \zeta_{mk} + 1\right) \nonumber \\
 	&&\!\!\!\!\! \!\!\!\!\! \!\!\!\!\! + P_d \sum_{m=1}^{M} \sum_{i \neq k}^{K} \eta_{mi} c_{mi}^2 \zeta_{mk} \left(\tau_p P_p \zeta_{mi} + 1 \right) \nonumber \\ 
	&&\!\!\!\!\!\! \!\!\!\!\! \!\!\!\!\! =P_d \sum_{m=1}^{M} \eta_{mk} \rho_{mk}^2  + P_d \sum_{m=1}^{M} \sum_{i = 1}^{K} \eta_{mi} \rho_{mi} \zeta_{mk}.\nonumber
\end{eqnarray}	
Furthermore, $\rho_{mk}$ in (\ref{eqn:avg_power_kth_Us}) can be derived as
\begin{eqnarray}\label{eqn:identity_2_1}
	\rho_{mk} &=& \E{\left|\hat{h}_{mk}\right|^2} = c_{mk}^2 \E{\left|y_{pmk}\right|^2}  \\
	&=& c_{mk} \E{\left(\sqrt{\tau_p P_p} h_{mk}^* + n_m\right) h_{mk}}= \sqrt{\tau_p P_p } c_{mk} \zeta_{mk}. \nonumber
\end{eqnarray}

\section{Derivation of SINR in \eqref{eqn:SINR_kth_US_analysis}} \label{app:Appendix3}
The expectation   in   numerator of \eqref{eqn:SINR_kth_US} can be derived   as  
	\begin{eqnarray}\label{eqn:identity_2}
		 \E {\sum_{m=1}^{M} \eta_{mk}^{1/2} h_{mk} \hat{h}_{mk}^*}  &=& \sum_{m=1}^{M} \eta_{mk}^{1/2} \E{\left(\hat{h}_{mk} + \epsilon_{mk} \right) \hat{h}_{mk}^* } \nonumber \\
		 &&\!\!\!\!\!\!\!\!\!\!\!\!\!\!\!\!\!\!\!\!\!\!\!\!\!\!\!\!\!\!\!\!\!\!\!\!\!\!\!\!\!\!\!\!\!\!\!\!\!  =  \sum_{m=1}^{M} \eta_{mk}^{1/2} \E{\hat{h}_{mk} \hat{h}_{mk}^* } =  \sum_{m=1}^{M} \eta_{mk}^{1/2} \rho_{mk},	
	\end{eqnarray}
where $\epsilon_{mk}$ is the error of  MMSE estimate of $h_{mk}$ such that  $h_{mk} = \hat{h}_{mk}+\epsilon_{mk} $.   
Then, the variance  in \eqref{eqn:SINR_kth_US} is derived as 
	\begin{eqnarray}\label{eqn:identity_4}
	 && \!\!\!\!\!\!\!\!\!\! \Var{\sum_{m=1}^{M} \eta_{mk}^{1/2} h_{mk} \hat{h}_{mk}^*} \nonumber \\
	&&
	= \sum_{m=1}^{M}  \eta_{mk} \left( \E{\left|\left( \hat{h}_{mk} +  \epsilon_{mk}  \right)  \hat{h}_{mk}^*\right|^2}   - \rho_{mk}^2 \right) \nonumber\\
	&& 
	=   \sum_{m=1}^{M}   \eta_{mk}  \left( 2\rho_{mk}^2  +  \rho_{mk} \left(\zeta_{mk}  -  \rho_{mk} \right)  -  \rho_{mk}^2 \right) \nonumber \\
	&&
	=  \sum_{m=1}^{M}  \eta_{mk} \rho_{mk} \zeta_{mk}. 
	\end{eqnarray}
The expectation in denominator of \eqref{eqn:SINR_kth_US} can be computed as
	\begin{eqnarray}\label{eqn:identity_5}
	&&\!\!\!\!\!\!\!\!\!\! \!\!\!\!\!\!\!\E {\left|\sum_{i \neq k}^{K}{\sum_{m=1}^{M} \eta_{mi}^{1/2} h_{mk} \hat{h}_{mi}^* } \right|^2 }  \nonumber \\
	&&
	\stackrel{(a)}{=}  \sum_{i \neq k}^{K} \sum_{m=1}^{M} \eta_{mi} c_{mi}^2 \E{\left| h_{mk} y_{pmi} \right|^2 } \nonumber\\
	&&
	\stackrel{(b)}{=}  \sum_{i \neq k}^{K}  \sum_{m=1}^{M}  \eta_{mi} c_{mi}^2  \zeta_{mk}  \left( \tau_p P_p \zeta_{mi}  + 1  \right)  \nonumber\\
	&&= \sum_{i \neq k}^{K}  \sum_{m=1}^{M}  \eta_{mi} \rho_{mi}  \zeta_{mk},
	\end{eqnarray}
where the steps $(a)$ and  $(b)$ are written by substituting \eqref{eqn:pilot_estimate} and \eqref{eqn:estimate_of_h_mk} into \eqref{eqn:estimate_of_h_mk} and \eqref{eqn:SINR_kth_US}, respectively.

\section{Derivation of effective SINR in \eqref{eqn:rate_DL_up_anlysis}} \label{app:Appendix4}

Next, $	\E{|\epsilon_{kk}^a|^2} $   in \eqref{eqn:rate_DL_up_full} is computed by substituting \eqref{eqn:akk_hat} and \eqref{eqn:Rx_DL_k} into the expectation of the denominator in \eqref{eqn:rate_DL_up_full} as  
	\begin{eqnarray}\label{eqn:epsilon_kk_a}
	\E{|\epsilon_{kk}^a|^2} &=& \E{|a_{kk} - \hat{a}_{kk}|^2} \nonumber \\ 
	&&\!\!\!\!\!\!\!\!\!\!\!\!\!\!\!\!\!\!\!\!\!\!\!\!\!\!\!\!
	= \E{\left|\frac{a_{kk} - \sum_{m=1}^{M} \eta_{mk}^{1/2} \rho_{mk} - \sqrt{\tau_{p,d} P_{p,d}} v_{kk} n_{pk,d} }{\tau_{p,d} P_{p,d} v_{kk} +1 } \right|^2 }  \nonumber \\
	&&\!\!\!\!\!\!\!\!\!\!\!\!\!\!\!\!\!\!\!\!\!\!\!\!\!\!\!\!
	= \frac{\E{\left|a_{kk} - \E{a_{kk}} - \sqrt{\tau_{p,d} P_{p,d}} v_{kk} n_{pk,d} \right|^2 }}{\left(\tau_{p,d} P_{p,d} v_{kk} +1\right)^2}  \nonumber \\
	&&\!\!\!\!\!\!\!\!\!\!\!\!\!\!\!\!\!\!\!\!\!\!\!\!\!\!\!\!
	= \frac{\Var{a_{kk}} +  \tau_{p,d} P_{p,d} v_{kk}^2}{\left(\tau_{p,d} P_{p,d} v_{kk} +1\right)^2}  
	= \frac{ v_{kk} }{\left(\tau_{p,d} P_{p,d} v_{kk} +1\right)}.
	\end{eqnarray}
The second expectation term in the denominator of \eqref{eqn:rate_DL_up_full} can be derived as
	\begin{eqnarray}\label{eqn:E_a_ki}
	\E{|a_{ki} |^2} &=& \E{\left|\sum_{m=1}^{M} \eta_{mi}^{1/2} \hat{h}_{mk} \hat{h}_{mi}^* \right|^2 }  \\
	&& \!\!\!\!\! \!\!\!\!\!+ \E{\left|\sum_{m=1}^{M} \eta_{mi}^{1/2} \epsilon_{mk} \hat{h}_{mi}^* \right|^2 } 
	= \sum_{m=1}^{M} \eta_{mi} \zeta_{mk} \rho_{mi}. \nonumber
	\end{eqnarray} 
 The expectation term in the numerator of  \eqref{eqn:rate_DL_up_full} is given by
\begin{eqnarray}\label{eqn:E_a_kk_hat}
	\!\!\!\!\!\!\!\!\!\! \E{|\hat{a}_{kk}|^2} &=& \E{|a_{kk} - \epsilon_{kk}^a |^2} = \E{|a_{kk}|^2} - \E{|\epsilon_{kk}^a |^2}.
\end{eqnarray} 
By substituting (30) in \cite{Ngo2016} and \eqref{eqn:epsilon_kk_a} into \eqref{eqn:E_a_kk_hat}, we have
\begin{eqnarray}\label{eqn:E_a_kk_hat_1}
	\E{|\hat{a}_{kk}|^2} &=& \sum_{m=1}^{M} \sum_{m'=1}^{M} \eta_{mk}^{1/2} \eta_{m'k}^{1/2} \rho_{mk} \rho_{m'k} \nonumber  \\
	&& \!\!\!\!\! \!\!\!\!\!+ \sum_{m=1}^{M} \eta_{mk} \zeta_{mk} \rho_{mk} - \frac{v_{kk}}{\tau_{p,d} P_{p,d} v_{kk} +1 }.
\end{eqnarray} 
Then, by substituting  \eqref{eqn:epsilon_kk_a}, \eqref{eqn:E_a_kk_hat} and \eqref{eqn:E_a_kk_hat_1} into \eqref{eqn:rate_DL_up_full},   the DL rate with DL pilots can be derived  as shown in  \eqref{eqn:rate_DL_up_anlysis}.

\section{Derivation of asymptotic signal power} \label{app:Appendix5}

As $M\rightarrow \infty$, the received signal power at the $k$th user   can be given as
\begin{eqnarray} \label{eqn:Rx_powe_inf} 
\!\!\!\!\!\!\!\!\!\!\!\!\!\!\!P_{k,{\infty}} &=& |r_{k,\infty}|^2 = \lim_{M \rightarrow \infty } \left|r_k \right|^2 \nonumber \\
&=& \lim_{M \rightarrow \infty } \left|\sqrt{P_d} \sum_{m=1}^{M} \sum_{i = 1}^{K} \eta_{mi}^{1/2}  h_{mk} \hat{h}_{mi}^* q_i  \right|^2\!\!\!,
\end{eqnarray}	 
 where $r_k$ is given in \eqref{eqn:rearranged_kth_Rx_signal}. By  using  channel estimate in \eqref{eqn:estimate_of_h_mk}, and then, by invoking Tchebyshev's theorem \cite{Cramer1970},  we have
\begin{eqnarray}\label{eqn:Tchebyshev's_signal}
	&&\!\!\!\!\!\!\!\! \frac{1}{M}   \sum_{m=1}^{M}   \eta_{mi}^{1/2} h_{mk} \hat{h}_{mi}^* -  \frac{1}{M} \sqrt{\tau_p P_p} \sum_{m=1}^{M} \!\! \eta_{mi}^{1/2} c_{mi} \zeta_{mk} \boldsymbol{\phi}_{i}^T \boldsymbol{\phi}_{k}^*  \nonumber\\
	&&\qquad\qquad\underset{M \rightarrow \infty}{\rightarrow} 0.
\end{eqnarray}
Since we use orthogonal pilots with $\boldsymbol{\phi}_{i}^T \boldsymbol{\phi}_{k}^* = 0$ for $i \neq k$, the residual interference due to pilot contamination  disappears  from the received signal. 
Thus, we have 
\begin{eqnarray}\label{eqn:asymptotic_Rx_signal}
	\frac{r_k}{M} - \frac{\sqrt{\tau_p P_p P_d}}{M} \sum_{m=1}^{M} \eta_{mk}^{1/2} c_{mk} \zeta_{mk} q_k \underset{M \rightarrow \infty}{\rightarrow} 0.
\end{eqnarray}
Then, the  signal received at the $k$th user can be asymptotically approximated as 
\begin{eqnarray}\label{eqn:asymptotic_Rx_signal_k}
	r_{k,\infty} = \lim\limits_{M \rightarrow \infty } r_k \rightarrow \sqrt{\tau_p P_p P_d} \sum_{m=1}^{M} \eta_{mk}^{1/2} c_{mk} \zeta_{mk} q_k.
\end{eqnarray}
Thus, the received signal at the $k$th user  can be written when $M \rightarrow \infty$ as
	\begin{eqnarray}\label{eqn:asymptotic_Rx_signal_k1}
	r_{k,\infty} &=& \sqrt{\tau_p P_p P_d} \sum_{m=1}^{M} \eta_{mk}^{1/2} c_{mk} \zeta_{mk} q_k \nonumber \\
	&=& \sqrt{P_d} \sum_{m=1}^{M} \eta_{mk}^{1/2} \rho_{mk} q_k.
	\end{eqnarray}

\noindent
By substituting \eqref{eqn:asymptotic_Rx_signal_k1} into \eqref{eqn:Rx_powe_inf},  the power of the received signal can be derived as
	\begin{eqnarray}\label{eqn:asymptotic_Rx_power1}
	P_{k,{\infty}} &=& |r_{k,\infty}|^2 =  P_d \left(\sum_{m=1}^{M} \eta_{mk}^{1/2} \rho_{mk} \right)^2.
	\end{eqnarray}

\linespread{1.0}



\begin{IEEEbiography}
	[{\includegraphics[width=1in,height=1.25in,clip,keepaspectratio]{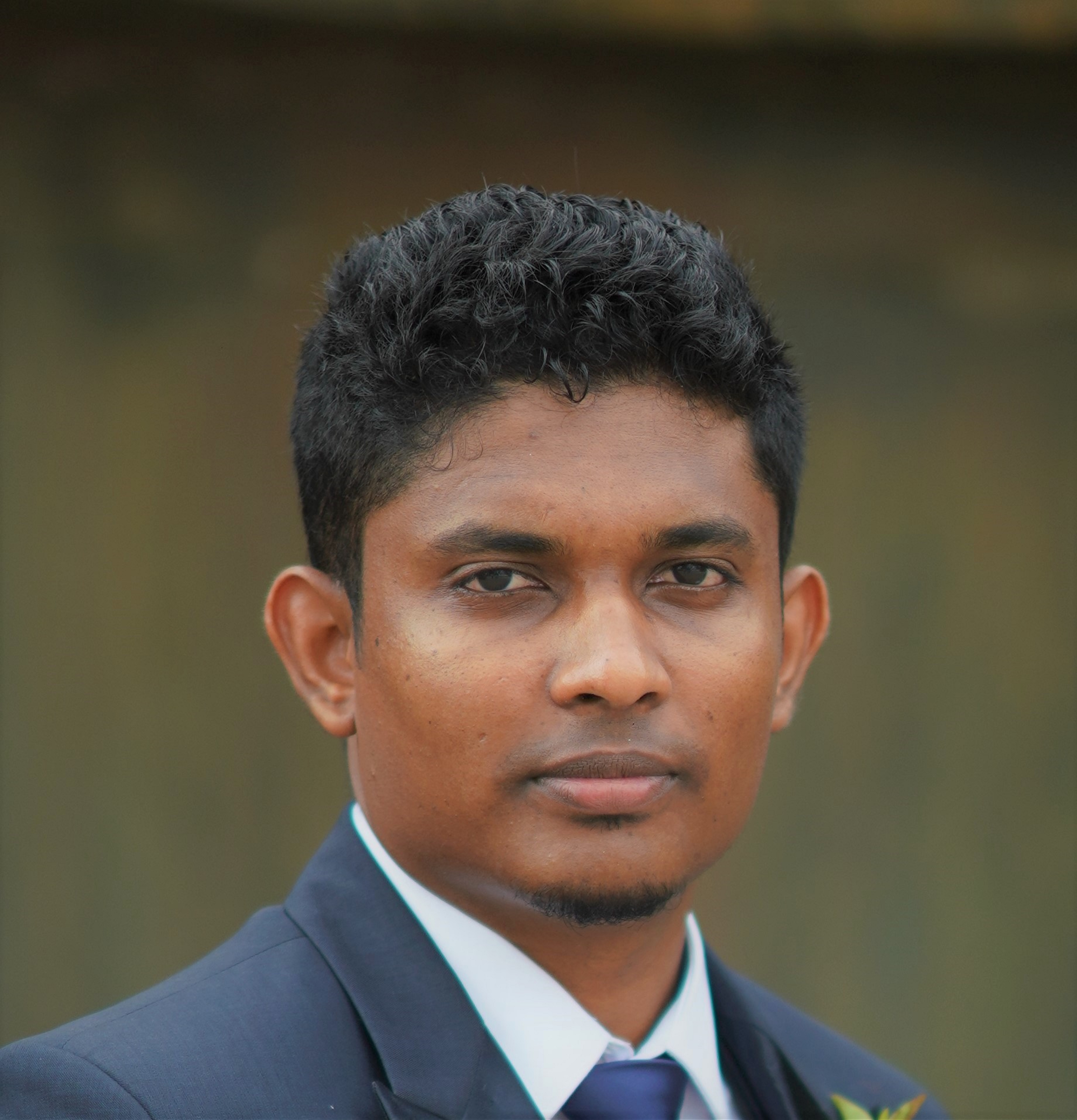}}]{Diluka Loku Galappaththige} (S'18)
	received the B. Sc. degree  (with first class
	Hons.)  from the Department of Electrical and Electronic Engineering, University of Peradeniya,  Sri Lanka, in  2017.  He is currently working towards the Ph.D. degree in the Department of Electrical and Computer Engineering, Southern Illinois University, Carbondale, IL, USA. 	
	His current research interests include   cell-free massive MIMO systems    and intelligent reflective surfaces.
\end{IEEEbiography}

\begin{IEEEbiography}
	[{\includegraphics[width=1in,height=1.25in,clip,keepaspectratio]{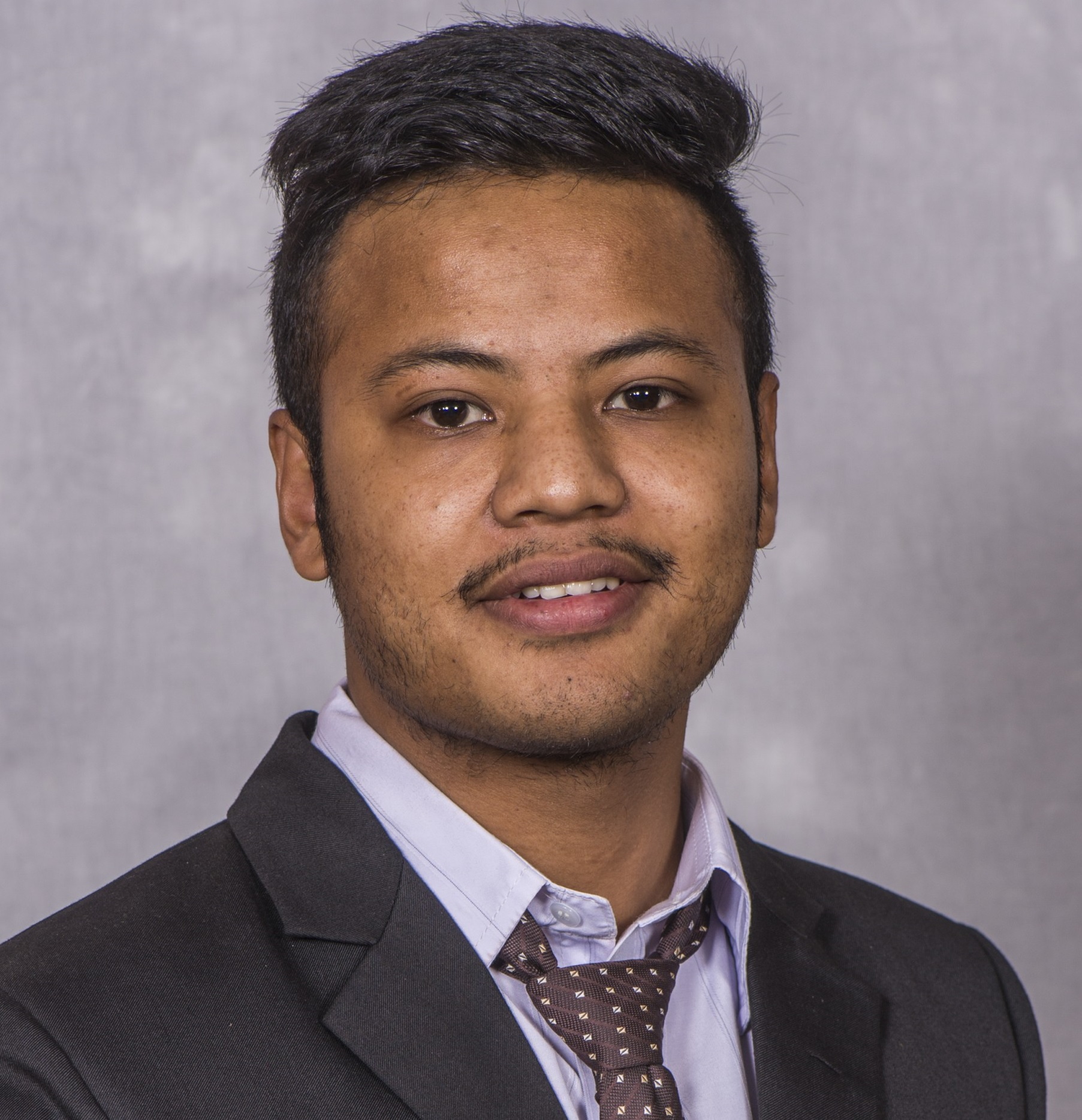}}]{Rajan Shrestha} 
	received the B.E. degree in electronics and communications engineering from Tribhuvan University, Nepal, in 2014 and the M.Sc. degree from the Department of Electrical and Computer Engineering, Southern Illinois University, Carbondale, IL, USA, in 2018. He is currently an RF Analyst with Global Wireless Solutions, Inc., Dulles, VA, USA. His current research interests include   wireless energy harvesting.  	
\end{IEEEbiography}

\begin{IEEEbiography}
	[{\includegraphics[width=1in,height=1.25in,clip,keepaspectratio]{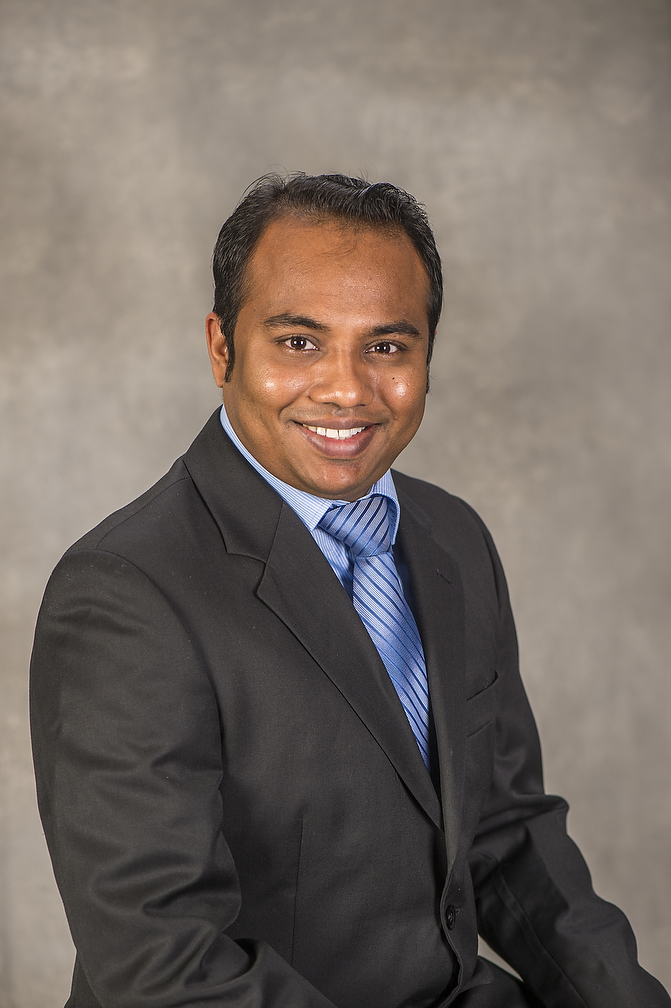}}]{Gayan        Amarasuriya Aruma Baduge} (S'09, M'13, SM'19) received the B.Sc. degree in  Engineering (with first class
	Hons.)  from the Department of Electronics and
	Telecommunications Engineering, University of
	Moratuwa, Moratuwa, Sri Lanka, in 2006, and the
	Ph.D. degree in Electrical Engineering from the
	Department of Electrical and Computer Engineering,
	University of Alberta, Edmonton, AB, Canada, in
	2013.  He was   a Postdoctoral Research Fellow
	with the Department of Electrical Engineering,
	Princeton University, Princeton, NJ, USA from 2014 to 2016.
	Currently, he is an assistant professor in     the Department of Electrical and Computer   Engineering  in Southern Illinois University, IL, USA.	
 	He is an Associate Editor for IEEE Communications Letters, IEEE Wireless Communications and IEEE Open Journal of the Communications Society.  
\end{IEEEbiography} 

\end{document}